\documentclass[useAMS,usenatbib]{mnras}

\usepackage{ amssymb }
\usepackage{mathptmx}
\usepackage{multirow}
\usepackage{psfrag}
\usepackage{pspicture}
\usepackage{graphicx}
\usepackage{amsmath}
\usepackage{color}
\usepackage{url}
\usepackage{caption}
\usepackage{subcaption}
\usepackage{grffile}

\title[A large narrow band H$\alpha$ survey at $z\sim0.2$]{A large narrow band H$\alpha$ survey at $z\sim0.2$: the bright end of the luminosity function, cosmic variance and clustering across cosmic time}
\author[A. Stroe et al.]{Andra Stroe$^{1}$\thanks{E-mail: astroe@strw.leidenuniv.nl} and David Sobral$^{1,2,3}$\thanks{VENI/IF Fellow}\\
$^{1}$Leiden Observatory, Leiden University, P.O.\ Box 9513, NL-2300 RA Leiden, The Netherlands\\
$^{2}$Instituto de Astro\'{\i}sica e Ci\^{e}ncias do Espa\c{c}o, Universidade de Lisboa, Observat\'{o}rio Astron\'{o}mico de Lisboa, Tapada da Ajuda, 1359-018, Lisbon, Portugal\\ 
$^{3}$Departamento de F\'{i}sica, Faculdade de Ci\^{e}ncias, Universidade de Lisboa, Edif\'{i}cio C8, Campo Grande, 1748-016, Lisbon, Portugal\\ 
}
\begin{document}
\maketitle
\begin{abstract}
We carried out the largest ($>3.5\times10^5$ Mpc$^3$, 26 deg$^2$) H$\alpha$ narrow band survey to date at $z\sim0.2$ in the SA22, W2 and XMMLSS extragalactic fields. Our survey covers a large enough volume to overcome cosmic variance and to sample bright and rare H$\alpha$ emitters up to an observed luminosity of $\sim10^{42.4}$ erg s$^{-1}$, equivalent to $\sim11 M_\odot$ yr$^{-1}$. Using our sample of $220$ sources brighter than $>10^{41.4}$ erg s$^{-1}$ ($>1 M_\odot$ yr$^{-1}$), we derive H$\alpha$ luminosity functions, which are well described by a Schechter function with $\phi^* = 10^{-2.85\pm0.03}$ Mpc$^{-3}$ and $L^*_\mathrm{H\alpha} = 10^{41.71\pm0.02}$ erg s$^{-1}$ (with a fixed faint end slope $\alpha=-1.35$). We find that surveys probing smaller volumes ($\sim3\times10^4$ Mpc$^3$) are heavily affected by cosmic variance, which can lead to errors of over $100$ per cent in the characteristic density and luminosity of the H$\alpha$ luminosity function. We derive a star formation rate density of $\rho_\mathrm{SFRD} = 0.0094\pm0.0008$ $M_\odot$ yr$^{-1}$, in agreement with the redshift-dependent H$\alpha$ parametrisation from \citet{2013MNRAS.428.1128S}. The two-point correlation function is described by a single power law $\omega(\theta) = (0.159\pm0.012) \theta^{(-0.75\pm0.05)}$, corresponding to a clustering length of $r_0 = 3.3\pm0.8$ Mpc/h. We find that the most luminous H$\alpha$ emitters at $z\sim0.2$ are more strongly clustered than the relatively fainter ones. The $L^*_\mathrm{H\alpha}$ H$\alpha$ emitters at $z\sim0.2$ in our sample reside in $\sim10^{12.5-13.5}$ $M_\odot$ dark matter haloes. This implies that the most star forming galaxies always reside in relatively massive haloes or group-like environments and that the typical host halo mass of star-forming galaxies is independent of redshift if scaled by $L_\mathrm{H\alpha}/L^*_\mathrm{H\alpha}(z)$, as proposed by \citet{2010MNRAS.404.1551S}.
\end{abstract}
\begin{keywords}
galaxies: luminosity function, mass function, galaxies: evolution, galaxies: formation, cosmology: large-scale structure of Universe
\vspace{-15pt}
\end{keywords}

\section{Introduction}\label{sec:intro}
The star formation (SF) activity in the Universe was significantly higher in the past, reaching a peak $\sim10-11$ Gyrs ago \citep[$z\sim2-3$, e.g.][]{1996ApJ...460L...1L, 2011ApJ...730...61K, 2011ApJ...737...90B,2013MNRAS.433.2764G, 2013MNRAS.428.1128S, 2015ApJ...803...34B}, and with the typical star formation rate (SFR) of galaxies (SFR$^*$) at $z\sim2$ being a factor $\sim10$ times higher than at $z = 0$ \citep{2014MNRAS.437.3516S}. However, the understanding of how and through which physical mechanisms the typical SFRs of galaxies have declined over the last $11$ Gyrs is still poor.

In order to study SF across cosmic time, a number of tracers can be used. Ultra violet (UV) data can be used to trace radiation coming from massive, short-lived stars. Dust heated by the UV radiation emits in the far infra-red (FIR). The radiation from the massive stars also ionises the surrounding gas and leads to numerous recombination lines such as H$\alpha$ ($6563${\AA}) and [OII] ($3727${\AA}). Radio observations can be used to trace emission from super nova remnants. However, it is not trivial to combine these SF indicators, given they trace different phases of SF (averaged on short, $\sim10$ Myr, or long, $\sim100$ Myr, timescales, dust obscured, etc.), with different selection functions. Some selections are significantly biased: UV-selected samples miss dusty/metal enriched star forming galaxies, while the FIR exclusively selects dusty star-forming regions. Therefore, one of the main challenges in obtaining a complete picture of the SF evolution is the direct comparison of equally selected large samples of SF galaxies at a range of redshifts. Samples at high redshift tend to be obtained with a completely different selection than those at lower redshift, which can result in misinterpreted evolutionary trends which are more likely connected with the different selections at different redshifts than the actual evolution of galaxies across time \citep[e.g.][]{2013MNRAS.430.1158S}.

\begin{table}
\begin{center}
\caption{Area and volumes covered by the narrow band observations. Only the common area between the two filters is listed. The same area is used to calculate the co-moving volume.}
\vspace{-5pt}
\begin{tabular}{l c c c c}
\hline\hline
Field & No pointings & Area & z & Volume \\
	& 	& deg$^2$ & &  $10^4$ Mp$c^3$ \\
\hline
\multirow{2}{*}{SA22} & \multirow{2}{*}{$24$} & \multirow{2}{*}{$6.1$} & $0.19$ & $7.5$ \\
	& 	& 	& $0.22$ & $9.8$ \\ \hline
\multirow{2}{*}{W2} & \multirow{2}{*}{$12$} & \multirow{2}{*}{$3.6$}  & $0.19$ & $4.4$ \\
	& 	& 	& $0.22$ & $5.7$ \\ \hline
\multirow{2}{*}{XMMLSS} & \multirow{2}{*}{$13$} & \multirow{2}{*}{$3.1$} &  $0.19$ & $3.9$ \\
	& 	& 	& $0.22$ & $5.0$ \\ \hline
Total & $49\times2$ & $12.8\times2$ &  & $36.3$ \\
\hline
\end{tabular}
\vspace{-10pt}
\label{tab:nbobs}
\end{center}
\end{table}

An effective way of overcoming such limitations is by using a single technique and a single SF indicator up to the peak of the star formation activity. This can be achieved by tracing the H$\alpha$ emission line, which is one of the most sensitive and well-calibrated SF traces and also benefits from low intrinsic dust extinction within the host galaxy (when compared to e.g. UV). H$\alpha$ surveys performed using the narrow-band (NB) technique can provide clean, large and complete samples SF galaxies (c.f. Oteo et al. 2015). 

A successful example of the NB technique put into practice is the High Redshift Emission Line Survey \citep[HiZELS,][]{2008MNRAS.388.1473G, 2010arXiv1003.5183B, 2013MNRAS.428.1128S}, but also see the pioneering works of \citet{1995MNRAS.273..513B}, \citet{2000A&A...362....9M}, \citet{2004A&A...428..793K}, \citet{2007ApJ...657..738L} and \citet{2008ApJS..175..128S}. At $z\sim1-2$, the volumes probed by HiZELS over a number of different fields ($\sim5-10$ deg$^2$) virtually overcome cosmic variance \citep{2015arXiv150206602S}. However, at $z < 0.4$, the volumes probed over $1-2$ deg$^2$ areas are only a minor fraction of those at high-redshift. Indeed, the samples at low redshift are greatly limited by cosmic variance, and even the widest surveys \citep[e.g.][Cosmological Evolution Survey (COSMOS)]{2008ApJS..175..128S} struggle to reach the characteristic H$\alpha$ luminosity ($L^*_\mathrm{H\alpha}$). An additional limitation is saturation, which means missing the luminous population of H$\alpha$ emitters \citep[with $>1-3$ M$_\odot$ yr$^{-1}$, for discussion of this effect see][]{2014MNRAS.438.1377S}. This can lead to an underestimation of H$\alpha$ luminosity function (LF) bright end and an exaggeration of the evolution of $L^*_\mathrm{H\alpha}$ from high to low redshift.

The combination of all these issues and the different selection techniques applied by each study makes it extremely hard to fairly compare between $z < 0.4$ and $z > 1$ samples when based on the same surveys. While it is possible to use other samples at lower redshift \citep[e.g. spectroscopic selection,][]{2013MNRAS.433.2764G}, the importance of using the same selection in order to obtain clean and clear evolutionary trends cannot be stressed enough: without the guarantee of a unique selection, any evolutionary trends become hard/impossible to understand and interpret, limiting our understanding. 

In order to overcome the current shortcomings we clearly require a large H$\alpha$ survey at lower redshifts which can be directly matched to higher redshift. In this paper we present a large survey at $z \sim 0.2$, covering a similar co-moving volume ($3.5\times10^5$ Mp$c^3$, spread over $3$ independent fields to overcome cosmic variance) and complete down to similar luminosity limits relative to $L^*_\mathrm{H\alpha}$ as surveys at $z>1$. The structure of the paper is as follows: in \S\ref{sec:obs-reduction} we present the observations and the reduction of the narrow-band data, while in \S\ref{sec:samples} we show the selection of the H$\alpha$ emitters. \S\ref{sec:LHA} deals with the $z\sim0.2$ H$\alpha$ luminosity function and \S\ref{sec:clustering} the clustering of bright H$\alpha$ sources and the implications of our results for the cosmic SF evolution are presented. We present concluding remarks in \S\ref{sec:conclusion}. 

At the two redshifts probed, $z\sim0.19$ and $0.22$, $1$ arcsec covers a physical scale of $3.2$~kpc and $3.6$~kpc, respectively. The luminosity distance is $d_\mathrm{L} \approx 940$ Mpc at $z\sim0.19$ and $\approx 1110$ Mpc at $z=0.22$. All coordinates are in the J2000 coordinate system. We use the \citet{2003PASP..115..763C} initial mass function (IMF) throughout the paper, and results from other studies are also converted to this IMF.

\section{Observations \& Data Reduction}
\label{sec:obs-reduction}

We obtain NB data tracing H$\alpha$ at $z\sim0.19$ and $\sim0.22$ in three well studied extragalactic fields located at high Galactic latitude. W2 is part of the Canada-France-Hawaii Telescope Legacy Survey (CFHTLS) $155$ deg$^2$, wide and shallow survey \citep{2012AJ....143...38G}, aimed at studying the large scale structure and matter distribution using weak lensing and galaxy distribution.  SA22 is part of the W4 field in CFHLS and multiwavelength data has been compiled by \citet{2014MNRAS.440.2375M} and \citet{2015arXiv150206602S}. The XMM Large Scale Structure Survey \citep[XMMLSS,][]{2004JCAP...09..011P} is aimed at mapping large scale structures through clusters and groups of galaxies.

\begin{table}
\begin{center} 
\caption{Typical $3\sigma$ limiting magnitudes for the three fields (including the standard spread in values), for each filter. The depth for each pointing (and within each CCD of out of the four WFC CCDs) varies across the fields over the ranges reported in the third and last column.}
\vspace{-5pt}
\begin{tabular}{l c c}
\hline\hline
Field & Filter & $3\sigma$ \\
	& 	 & mag \\
\hline
\multirow{2}{*}{SA22} & NB1 & $17.5^{+0.4}_{-0.3}$
  \\
	& NB2 & $17.4^{+0.4}_{-0.3}$  \\ \hline
\multirow{2}{*}{W2} & NB1 &  $16.8^{+1.5}_{-0.6}$ \\
& NB2 &  $16.7^{+0.7}_{-0.4}$  \\ \hline
\multirow{2}{*}{XMMLSS} & NB1 & $17.7^{+0.4}_{-0.3}$ \\
& NB2 & $17.5^{+0.5}_{-0.3}$  \\ \hline
\hline
\end{tabular}
\vspace{-10pt}
\label{tab:limmag}
\end{center}
\end{table}

\begin{figure*}
\begin{center}
\includegraphics[trim=0cm 0cm 0cm 0cm, width=0.995\textwidth]{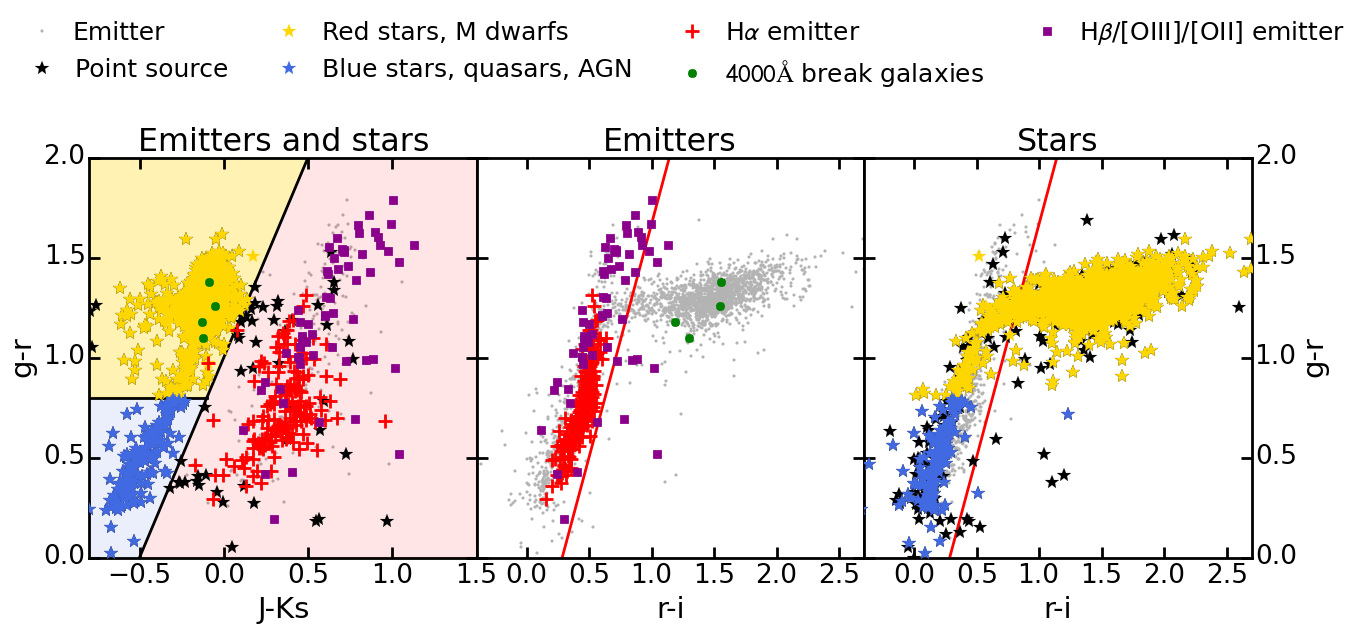}
\end{center}
\vspace{-10pt}
\caption{Colour-colour plots for the SA22, W2 and XMMLSS fields, mainly used to remove stars. The first plot shows $g-r$ versus $J-Ks$ while the middle and last plots show $g-r$ versus $r-i$. We first separate stars and emitters using the $g-r$ versus $J-Ks$, and the apply an extra cut using the optical colours to further remove stars with absorption features in one of the filter. The solid red and black lines display the colour cuts used to select point-like objects. H$\alpha$ emitters are plotted in red crosses, while point-like sources are plotted as stars. $4000${\AA} break galaxies are plotted in green crosses and high redshift sources in purple crosses.}
\vspace{-10pt}
\label{fig:colcol}
\end{figure*}

\subsection{Narrow band H$\alpha$ observations}

We obtained narrow band data using the NOVA782HA and NOVA804HA \citep{2014MNRAS.438.1377S,Stroe2015,2015MNRAS.450..630S} filters on the Wide
Wide Field Camera (WFC)\footnote{\url{http://www.ing.iac.es/engineering/detectors/ultra_wfc.htm}} mounted on the Isaac Newton Telescope (INT, I13BN008, PI Sobral) \footnote{\url{http://www.ing.iac.es/Astronomy/telescopes/int/}}. For brevity, we label the filters as NB1 (NOVA782HA) and NB2 (NOVA804HA). Given the central wavelengths of the filters are $7852.4${\AA} and $8036.15${\AA}, with a full-width-half-maximum (FWHM) of $110${\AA}, the two filters trace H$\alpha$ emission in the $z=0.1865-0.2025$ and $z=0.2170-0.2330$ redshift ranges. Note that given the large field of view of WFC, a slight blue shift in the filter central wavelength is expected at large off-axis distances. However, given the WFC focal ratio ($f/3.29$), this effect is expected to be very low \citep[a few per cent][]{2001ApJ...563..611B}. \citet{2015MNRAS.450..630S} and \citet{Stroe2015} characterised the filters with spectroscopy from the Keck and Willam Herschel Telescopes with sources located both towards and away from the pointing centre and found that the redshift distribution of H$\alpha$ emitters matches that expected from the filter profile, without any noticeable offset.

Observations were conducted in five bright nights, between 22 and 26 of October 2013, under $\sim1$ arcsec seeing conditions. A five-position dither pattern was employed for the individual exposures (of $600$ s each) to cover the spacings between the four WFC CCDs. Forty-nine individual pointings (of $\sim0.3$ deg$^2$ each with WFC) split between the three fields (SA22, W2 and XMMLSS) cover an area of almost $13$ deg$^2$ at each of the two redshifts (thus an effective area of $\sim26$ deg$^2$ combined), tracing a total co-moving volume of about $3.63\times 10^5$ Mpc$^3$. The overlap with the multiwavelength data extends to about $10$ deg$^2$ per redshift.

\subsection{Narrow band data reduction}\label{sec:obs:NB}
We reduce the data using the \textsc{python} based pipeline described in \citet{2014MNRAS.438.1377S}. In short, we median combine the sky flats and biases and use the stacks to correct the science data. After detecting sources using the {\sc SExtractor} package \citep{1996A&AS..117..393B} and masking them in each science exposure, we median combine the exposures to obtain a `super-flat'. We divide the data through the `super-flat' to correct for `fringing'. We then use {\sc SCAMP} \citep{2006ASPC..351..112B} to find astrometric solutions for the science exposures. The exposures are normalised to the same zero-point (ZP) by comparison to the red magnitude in the fourth United States Naval Observatory (USNO) Catalog \citep[UCAC4;][]{2013AJ....145...44Z}. We combine the processed data into final stacked images using {\sc SWarp} \citep{2002ASPC..281..228B}. We photometrically calibrate our data against the i band magnitude from the Sloan Digital Sky Survey (SDSS) Data Release 9 \citep[SDSS DR9][]{2012ApJS..203...21A}, which covers all our fields (SA22, W2 and XMMLSS). We extract magnitudes within $5$ arcsec apertures using {\sc SExtractor} \citep{1996A&AS..117..393B}. This corresponds to a physical diameter of $\sim18$ kpc at $\sim0.2$ redshift. 

We calculate $3\sigma$ limiting magnitudes using the RMS noise reported by {\sc SExtractor} (see Table~\ref{tab:limmag}). The depth of the observations varies across the pointings and even between the different chips of the WFC. Hence, we calculate the RMS noise individually for each CCDs, for each pointing, across the three fields.

We apply the NB technique to select line emitters, using a NB filter tracing line emission within a narrow range in redshift, in combination with another NB or broad band (BB) filter used for the estimation of the continuum emission underlying the emission line. We use two NB filters to trace H$\alpha$ emission in two redshift ranges ($0.1865-0.2025$ and $0.2170-0.2330$). For each NB filter, we use the other NB filter to estimate the continuum BB emission. In this way, for line emitters, one NB filter captures the BB emission as well as the line emission, while the other NB filter only captures the stellar continuum emission. Our method is similar to that of \citet{2010ApJ...712L.189D}, who use twin NB filters for continuum subtraction. In further text, we use labels according to the filter which was used as NB filter in that particular case. Therefore, when we label with NB1, we refer to line-emitters in the $0.1865-0.2025$ redshift range, while NB2 refers to the $0.2170-0.2330$ range. The details of the selection method are laid down in \S\ref{sec:samples:selection}.

\begin{figure*}
\centering
\begin{subfigure}[b]{0.49\textwidth}
\includegraphics[trim=0cm 0cm 0cm 0cm, width=0.995\textwidth]{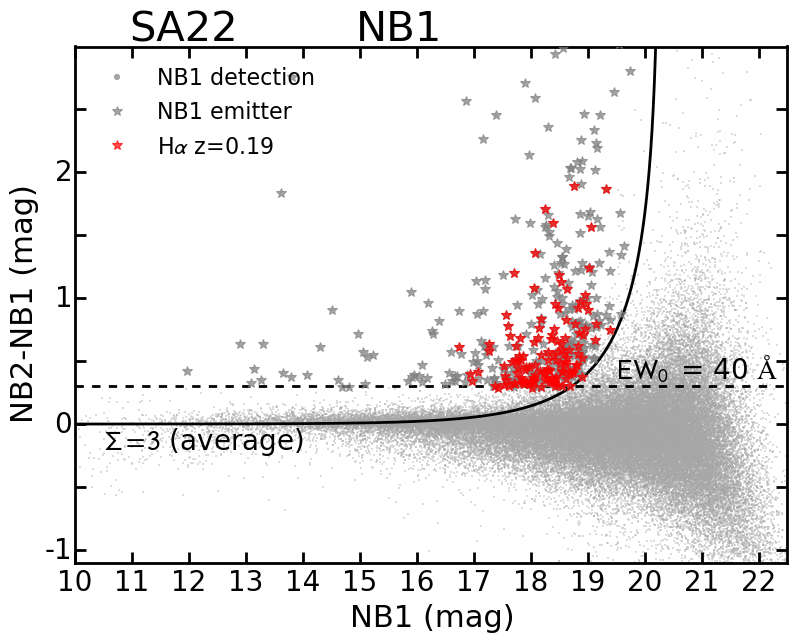}
\end{subfigure}
\hspace{5pt}
\begin{subfigure}[b]{0.49\textwidth}
\includegraphics[trim=0cm 0cm 0cm 0cm, width=0.995\textwidth ]{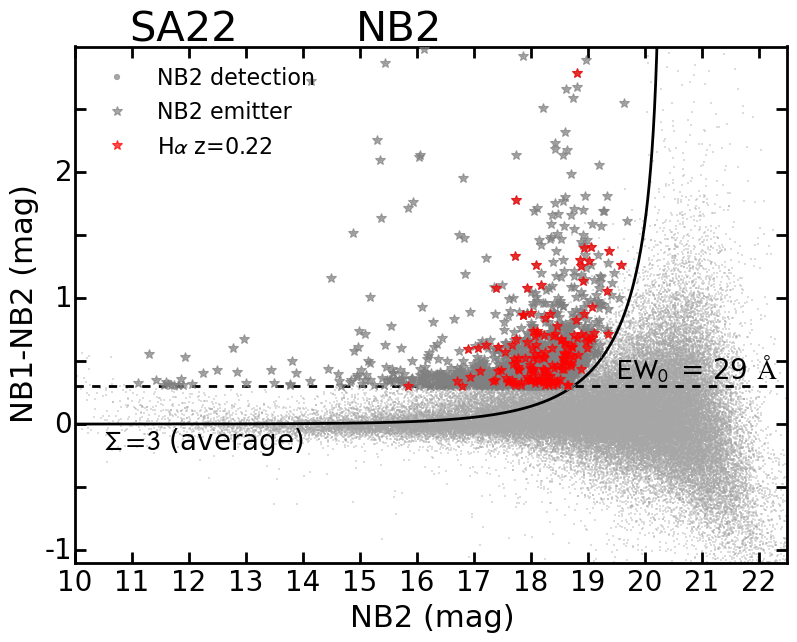}
\end{subfigure}
\begin{subfigure}[b]{0.49\textwidth}
\includegraphics[trim=0cm 0cm 0cm 0cm, width=0.995\textwidth ]{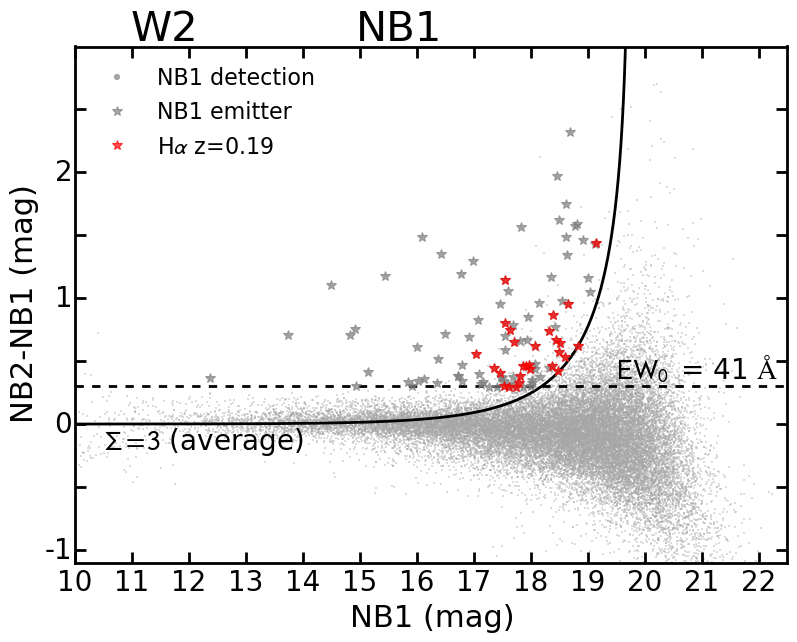}
\end{subfigure}
\hspace{5pt}
\begin{subfigure}[b]{0.49\textwidth}
\includegraphics[trim=0cm 0cm 0cm 0cm, width=0.995\textwidth ]{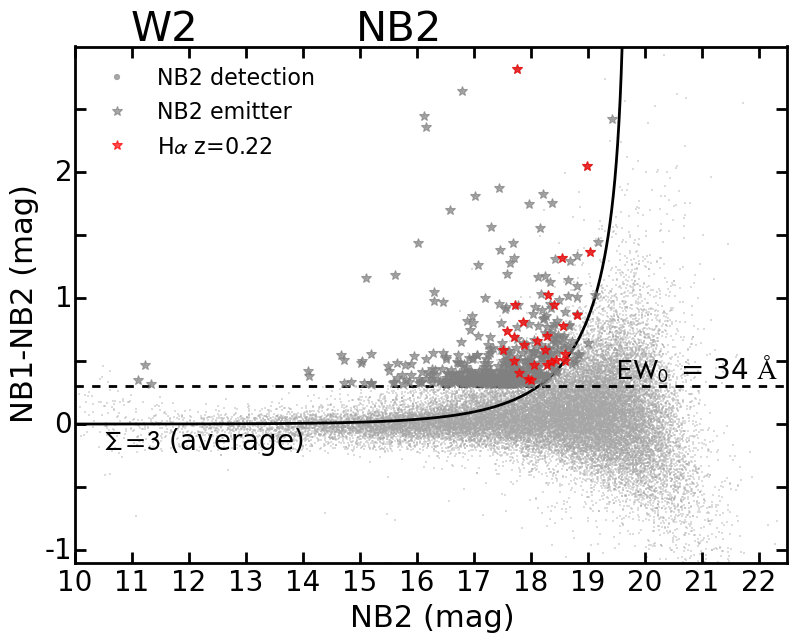}
\end{subfigure}
\begin{subfigure}[b]{0.49\textwidth}
\includegraphics[trim=0cm 0cm 0cm 0cm, width=0.995\textwidth ]{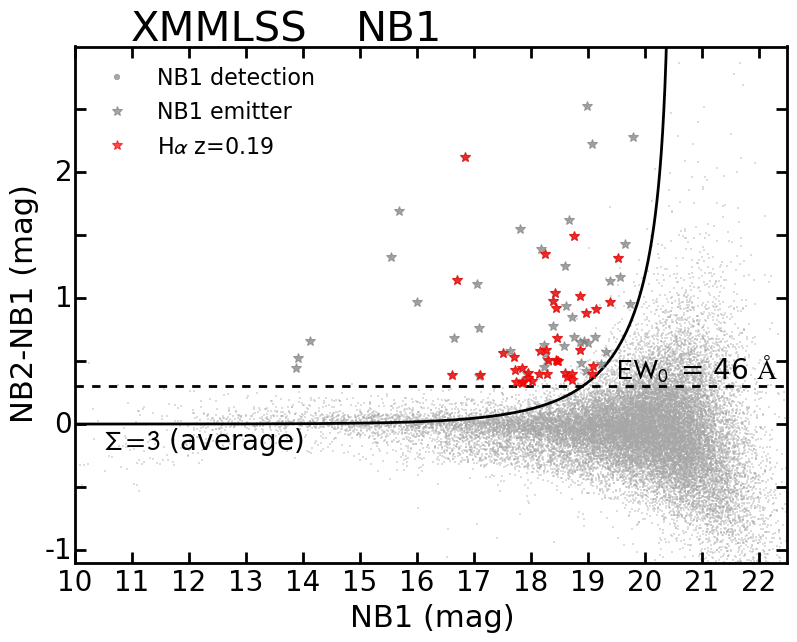}
\end{subfigure}
\hspace{5pt}
\begin{subfigure}[b]{0.49\textwidth}
\includegraphics[trim=0cm 0cm 0cm 0cm, width=0.995\textwidth ]{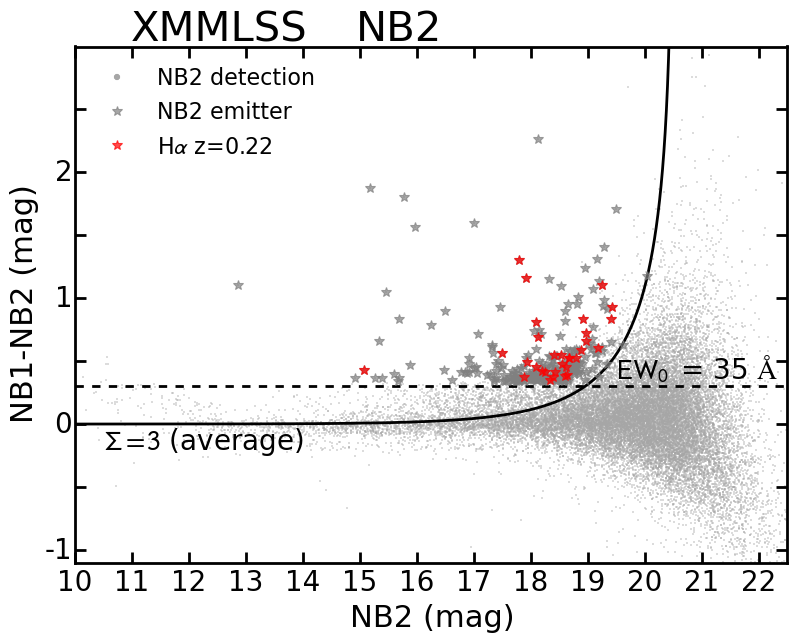}
\end{subfigure}
\caption{Colour-magnitude diagrams showing the excess as function of NB magnitude. The selection is performed separately for each CCD/pointing, field and NB filter, using the other NB filter for continuum estimation. Each panel is labelled with the corresponding field and the filter which is used as NB. The curves show average $3\Sigma$ colour significances for the average depth, as the RMS value varies between the pointings and CCDs. The horizontal dashed, black lines represent the intrinsic EW cuts. Note that we correct for incompleteness arising from our slightly different EW and colour significance cuts.}
\label{fig:colmag}
\end{figure*}

\subsection{Optical and IR data}\label{sec:obs:optIR}
In our analysis, we use the rich multi-wavelength optical and infra-red (IR) data available for the SA22, W2 and XMMLSS fields.

All three fields are part of the CFHTLS wide and shallow surveys (SA22, W2 and XMMLSS are in fields W4, W2 and W1). We make use of the $g$, $r$, $i$ and $z$ photometry \citep{2013MNRAS.433.2545E} and photometric redshifts \citep{2006A&A...457..841I} available through the CFHTLS T0007 release. 
 
We also employ near IR data in the $J$ and $Ks$ filters, down to magnitude $\sim21.2$ and $\sim20.0$ respectively, obtained as part of the Visible and Infrared Survey Telescope for Astronomy (VISTA) Hemisphere Survey (VHS, McMahon and the VHS Collaboration, 2012, in preparation). Where available, in the XMMLSS field, we preferentially use data from the VISTA Deep Extragalactic Observations (VIDEO) Survey \citep{2013MNRAS.428.1281J}, which is about $3.5$ magnitudes deeper than VISTA. We also make use of the IR photometric data taken in the SA22 field as part of the second data release of the UKIDSS Deep Extragalactic Survey \citep{2007astro.ph..3037W}, which reaches magnitudes $23.4$ and $22.8$ in the $J$ and $Ks$ bands, respectively, with a catalogue from \citet{2015arXiv150206602S}. 

We make use of the photometric and spectroscopic redshift compilation in the UKIDSS Ultra Deep Survey (part of XMMLSS) available as part of their 8th data release\footnote{\url{http://www.nottingham.ac.uk/astronomy/UDS/data/data.html}}, as well as other publicly available spectroscopy in the XMMLSS field \citep{2007A&A...474..473G, 2007ApJ...663...81P, 2007A&A...467...73T, 2013A&A...557A..81M}.

\section{Methods and selecting the H$\alpha$ samples}\label{sec:samples}

Once sources are detected in the NB images, we cross-match the NB catalogues with the optical and IR catalogues presented in Section \ref{sec:obs:optIR}, using a $1$ arcsec positional tolerance. Note that because the BB catalogues are deeper than our data by at least $2$ mag, we have $100$ per cent optical and IR coverage in the areas we have FOV overlap with all the multiwavelength data. We use each NB catalogue as base catalogue for the cross-match.

\begin{table}
\begin{center}
\caption{Number of line emitters and H$\alpha$ emitters selected in each field and filter. We also list the average limiting observed H$\alpha$ luminosity at $50$ per cent completeness and the equivalent SFR (using equation \ref{eq:SFR}).}
\vspace{-5pt}
\begin{tabular}{l c c c c c}
\hline\hline
Field & Filter & Emitters & H$\alpha$ emitters & $\log(L_\mathrm{H\alpha})$ & SFR \\
& & & &  (erg s$^{-1}$) & ($M_\odot$ yr$^{-1}$)\\
\hline
\multirow{2}{*}{SA22} & NB1 & $153$ & $59$ & $41.4$ & $1.1$ \\
	& NB2 & $238$ & $91$  & $41.4$ & $1.1$ \\ \hline
\multirow{2}{*}{W2} & NB1  & \phantom{0}$33$ & $13$ & $41.4$ & $1.1$\\
& NB2 & \phantom{0}$55$ & $15$ & $41.6$ & $1.7$ \\ \hline
\multirow{2}{*}{XMMLSS} & NB1 & \phantom{0}$51$ & $23$  & $41.1$ & $0.5$ \\
& NB2 & \phantom{0}$50$ & $19$  & $41.4$ & $1.1$\\ 
\hline
Total & both & $576$ & $220$ \\
\hline
\end{tabular}
\vspace{-10pt}
\label{tab:number}
\end{center}
\end{table}

\subsection{Star removal}\label{sec:samples:point}

As explained in \S\ref{sec:obs:NB}, we use the two NB filters to trace H$\alpha$ emission at two redshifts ranges ($0.1865-0.2025$ and $0.2170-0.2330$). However, given the wavelength coverage of the two adjacent filters our samples of line emitters is contaminated by stars \citep[see also][]{2014MNRAS.438.1377S}. Stars could mimic having an emission line if they have extremely red or a broad absorption feature, which would lead to a strong colour between the two NB filters. We expect the line emitters selected in the NB2 filter to be particularly contaminated with a population of (L, M) dwarf stars \citep{1991ApJS...77..417K, 1999ApJ...519..802K}. They will be selected as having excess in NB2 because their continuum has a broad absorption feature falling within the NB1 filter, leading to an underestimation of the continuum emission. The extremely red BB colours of these sources are also consistent with them being red dwarfs. 


We exclude stars using a colour-colour selection criterion using optical and IR colours based on \citet{2012MNRAS.420.1926S}, keeping in mind the distribution of sources in the colour-colour diagram. This is illustrated in Figure \ref{fig:colcol}.

Red stars are selected using:
\begin{equation}
\label{eq:redstars}
(g-r) > 2 (J-Ks) + 1 \quad \& \quad (g-r)>0.8 \quad  \& \quad (J-Ks) > -0.7
\end{equation}

We select dwarf stars via:
\begin{equation}
\label{eq:optstars}
(g-r) > ( 7/3 (r-i) - 2/3 ) \quad \& \quad (g-r)>1.0
\end{equation}

Optically blue stars and dwarf stars with absorption features are selected by:
\begin{equation}
\label{eq:bluestars}
(g-r) > 2 (J-Ks) + 1 \quad \& \quad (g-r)<0.8 
\end{equation}

We additionally use the `StarGal' parameter in the CFHTLS photometric redshift catalogue to select stars \citep{2006A&A...457..841I}, which categorises sources as point-like or extended objects.

Thus, in summary, we label sources as stars if:
\begin{itemize}
\item Source passes the red star selection criterion (equation \ref{eq:redstars}) or
\item Source passes the blue star selection criterion (equation \ref{eq:bluestars}) or
\item Source passes the dwarf star selection criterion (equation \ref{eq:optstars}) or
\item Source is classified as star by the CFHTLS `StarGal' parameter.
\end{itemize}

About $60-80$ per cent of the sources mimicking emission lines are marked as stars. Spectroscopic observations using NB1 and NB2 \citep[e.g][]{Stroe2015, 2015MNRAS.450..630S} confirm the presence of such stars. All the sources masked as stars are removed from catalogues such that they are not selected as line emitters. 

\subsection{Selection of line emitters}\label{sec:samples:selection}

We use the formalism developed by \citet{1995MNRAS.273..513B}, which is widely used in the literature \citep[e.g.][]{2008ApJS..175..128S, 2009MNRAS.398...75S, 2014MNRAS.438.1377S} to select large numbers of line emitters. We refer the interested readers to those papers for the details of the selection criteria. 

We select line emitters separately in each field and each NB filter. For brevity, in the following equations, we label the filter used to select emitters as NB, while we name the other NB filter, used to quantify the continuum emission, as BB filter. Note that we attempted the selection of line emitters using the i band filter, following \citet{2014MNRAS.438.1377S} and \citet{Stroe2015}. However, the relatively deep CFHTLS data becomes saturated at magnitude $17-18$ and would prevent the selection of bright line emitters. Therefore using each NB filter for continuum subtraction of the other represents the optimal strategy, enabling the selection of line emitters up to magnitude $10$. Using much deeper broad band $i$ data would allow us to probe down to fainter emitters, but our aim for the paper is to study the bright population. By comparison, the widest H$\alpha$ survey at $z\sim0.2$ to date, performed by \citet{2008ApJS..175..128S}, can only probe sources as bright as $\sim18$ mag, but excels at the faint end (going down to $24$ mag).

We select emitters in each NB filter based on their excess emission compared to the BB emission (quantified using the other NB filter). We first correct for any systematic colour offset between the two NB filters. Colour is defined here as the difference in magnitude between the filter used as NB and the filter used to measure broad band. We estimate a median offset of this colour, based on the scatter in the colours at non-saturated, but still bright NB magnitudes. We then apply this correction to the colour and the NB magnitude. However, because the filters are close in wavelength this correction is small ($0.02$ and $0.03$ mag, for NB1 and NB2 respectively). 

The excess emission is then quantified through the colour excess significance $\Sigma$, which is used to separate sources with real colour excess, compared to excess caused by random scatter \citep{2009MNRAS.398...75S,2012MNRAS.420.1926S}:
\begin{equation}
\label{eq:Sigma}
\Sigma = \frac{10^{-0.4\left(m_{BB}-m_{NB}\right)}}{10^{-0.4(ZP_{AB}-m_{NB})} \sqrt{\pi r^2 \left(\sigma^2_\mathrm{NB}+\sigma^2_\mathrm{BB}\right)}},
\end{equation}
where $ZP_{AB}$ is the magnitude system zero-point, $m_{NB}$ and $m_{BB}$ are the NB and BB magnitudes (where NB is the filter used for detection of line emitters and BB is the other NB filter used for quantifying the continuum emission), $r$ is the radius of the aperture in pixels and $\sigma_\mathrm{NB}$ and $\sigma_\mathrm{BB}$ are the rms noise levels. 

The NB or BB flux $f_{NB,BB}$ are calculated as:
\begin{equation}
\label{eq:flux_broad}
f_{NB,BB} = \frac{c}{\lambda^2_{NB,BB}} 10^{-0.4(m_{NB,BB}-ZP_{AB})},
\end{equation}
where $c$ is the speed of light, $\lambda_\mathrm{NB}$ and $\lambda_\mathrm{BB}$ are the central wavelengths of the two NB filters and $ZP_{AB}=48.574$ is the ZP of the AB magnitude system. The line flux is:
\begin{equation}
\label{eq:flux}
F_\mathrm{line} = \Delta\lambda_\mathrm{NB} (f_{NB}-f_{BB}).
\end{equation}
Note that the two filters are independent, hence there is no overlap in wavelength between NB1 and NB2. Therefore, if one filter captures line emission on top of the continuum, automatically the other NB filter picks up only continuum emission. Therefore, the line flux formula accounts for the fact the filter used as BB does not contain any line emission.

\begin{figure}
\begin{center}
\includegraphics[trim=0cm 0cm 0cm 0cm, width=0.495\textwidth]{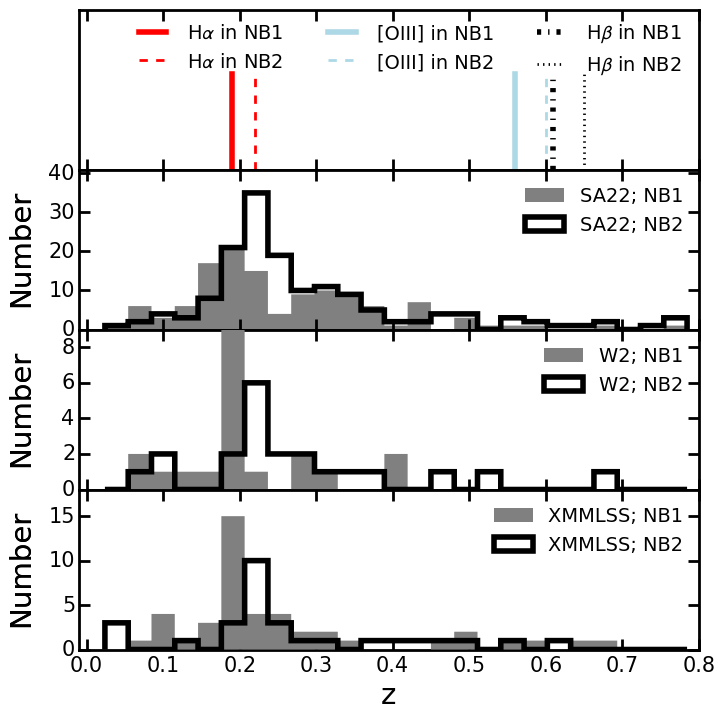}
\end{center}
\vspace{-10pt}
\caption{Photometric redshift distribution of line emitters for each field. Note the quality of the photometric redshifts varies between the fields. The top panel shows the main line we expect to capture with out two narrow band filters. The distribution contains clear peaks around $z\sim0.2$, indicating our sample is dominated by H$\alpha$ emitters, with little contamination from higher redshift emitters.}
\vspace{-10pt}
\label{fig:histz}
\end{figure}

We use the $\Sigma$ parameter in conjunction with an equivalent width ($EW$) cut, which ensures that we select only sources which have a ratio of the line to continuum flux larger than the scatter at bright magnitudes. The observed $EW$ is defined as:
\begin{equation}
\label{eq:EW}
\mathrm{EW} = \Delta\lambda_\mathrm{NB} \frac{f_{NB}-f_{BB}}{f_{BB}},
\end{equation}
where $\Delta\lambda_\mathrm{NB}=100$ {\AA} is the FWHM of the NB filters, while $f_{NB}$ and $f_{BB}$ are the NB and continuum fluxes. Note this formula is a simplified version of those presented in, e.g., \citet{1995MNRAS.273..513B} and \citet{2009MNRAS.398...75S}, because we do not expect our BB filter to contain any emission line flux.

In the restframe of the sources, the intrinsic $EW_0$ is:
\begin{equation}
\label{eq:EW0}
\mathrm{EW}_0 = \mathrm{EW}/\left(1+z\right).
\end{equation}

In conclusion, we select sources as emitters if:
\begin{itemize}
\item Their colour significance $\Sigma$ is higher than $3$ and
\item Their equivalent width is higher than $3\sigma$, where $\sigma$ is the colour excess (BB-NB) scatter at bright, but not saturated magnitudes.
\end{itemize}
The $\Sigma=3$ colour significance and the $3\sigma$ excess depend on the depth of the observations in each field (See Figure \ref{fig:colmag}). We choose to not impose a single, common cut, to follow the natural depth of the data, rather than cutting the sample at excessively high $EW$ and $\Sigma$. However, we note that we fully correct for the sources missed by our cuts, as explained in \S\ref{sec:LHA:completeness}.

\begin{figure}
\centering
\includegraphics[trim=0cm 0cm 0cm 0cm, width=0.445\textwidth]{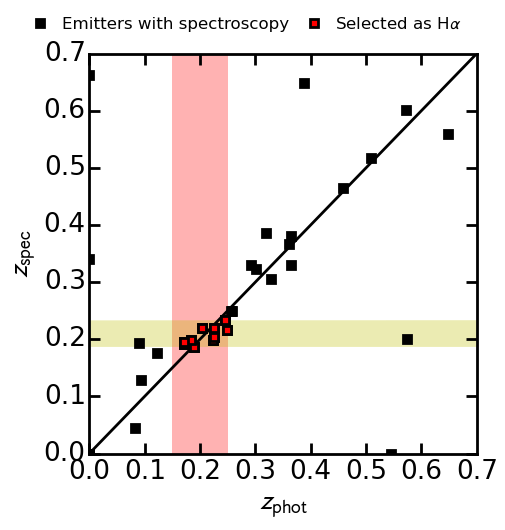}
\caption{Photometric versus spectroscopic redshift for sources selected as emitters. The shaded red area indicates sources which based on their photometric redshift were selected as H$\alpha$. The yellow shaded area indicates the redshift range captured by the filters.}
\label{fig:zdistrib}
\end{figure}

\subsection{Selection of H$\alpha$ candidates}\label{sec:samples:selectionHA}

The line emitter population is made of H$\alpha$ emitters at $z\sim0.2$, as well as higher redshift line emitters: H$\beta$ ($\lambda_\mathrm{rest}=4861$\;{\AA}), [O{\sc iii}]$\lambda\lambda4959,5007$ emitters at $z\sim0.61-0.65$ and [O{\sc ii}] ($\lambda_\mathrm{rest}=3727$\;{\AA}) emitters at $z\sim1.15$ (see Figure \ref{fig:histz}). Our sample could be contaminated by a population of $4000$\;{\AA} break galaxies at $z\sim0.8$. As shown in \citet{2014MNRAS.438.1377S}, at $\sim8000$\;{\AA} and lower line fluxes, the line emitter population is dominated by [O{\textsc II}]$\lambda3727$ emitters and $z\sim0.8$ $4000$\;{\AA} break galaxies. However, at high fluxes, the number of H$\alpha$ and H$\beta$/[O{\textsc III}] steeply rises, each amounting to about $50$ per cent of the line emitter population. Therefore, given the shallow depth of our survey, we are strongly biased against detecting high-redshift ($z>0.6$) sources. We expect the H$\alpha$ emitters to amount to about half of the emitter population. Figure~\ref{fig:histz}, presenting the photometric redshift distribution of the line emitters, confirms these findings. The steps we undertake to robustly separate the H$\alpha$ emitters from the other sources are described in the following paragraphs.

We first visually inspected all line emitter candidates to flag any spurious sources coming from noisy edge regions of the chips or from false detections within the haloes of bright sources. 

\begin{figure*}
\centering
\includegraphics[trim=0cm 0cm 0cm 0cm, width=0.995\textwidth]{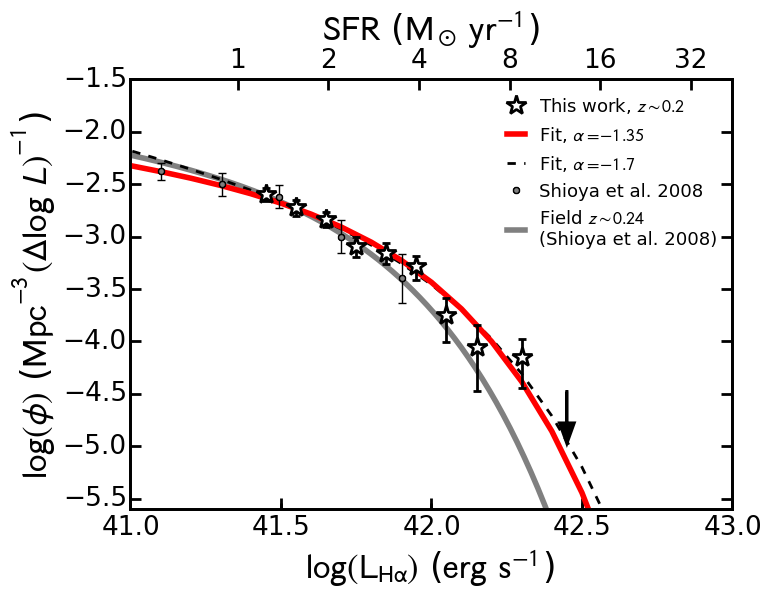}
\caption{The H$\alpha$ luminosity function at $z\sim0.2$ from our study and the best fit Schechter function. The $L_\mathrm{H\alpha}$ is not corrected for intrinsic dust attenuation. For comparison, the results from \citet{2008ApJS..175..128S} are also shown. Note the excellent agreement between the data in the overlapping luminosity range. However, our data probes brighter luminosities, enabling the first determination of the bright end of the H$\alpha$ luminosity function at $z\sim0.2$.}
\label{fig:bestLF}
\end{figure*}

\begin{table*}
\begin{center}
\caption{Best fit luminosity function at $z\sim0.2$ obtained from combining data in the three fields (SA22, W2 and XMMLSS) and two NB filters. Since our data is not very deep, but probes the bright-end really well, we fix the faint-end slope $\alpha$ at two values. For comparison, we also list the results and volumes probed from other studies at a similar redshift. Note that none of the $L^*_\mathrm{H\alpha}$ are corrected for H$\alpha$ extinction.}
\vspace{-5pt}
\begin{tabular}{l c c c c c }
\hline\hline
Source & $z$ & $V$ ($10^4$ Mpc$^3$) & $\alpha$ & $\log \phi^*$ (Mpc$^{-3}$) & $\log L^*_\mathrm{H\alpha}$ (erg s$^{-1}$) \\
\hline
\multirow{3}{*}{This study} & \multirow{2}{*}{$\sim0.2$}  &  \multirow{2}{*}{$36.3$}  &  $-1.35$ & $-2.85 \pm 0.03$ & $41.71 \pm 0.02$ \\
& & & $-1.70$ & $-3.06 \pm 0.04$ & $41.83 \pm 0.03$ \\ \hline
\citet{2008ApJS..175..128S} & $\sim0.24$ & $3.1$ & $-1.35^{+0.11}_{-0.13}$ & $-2.65^{+0.27}_{-0.38}$ & $41.54^{+0.38}_{-0.29}$ \\
\citet{2007ApJ...657..738L} & $\sim0.24$ & $0.5$ & $-1.70\pm0.10$ & $-2.98\pm0.40$ & $41.25\pm0.34$ 
\\
\citet{2013MNRAS.433..796D} & $\sim0.25$ & $1.2$ & $-1.03^{+0.17}_{-0.15}$ &  $-2.53^{+0.17}_{-0.21}$ & $40.83^{+0.19}_{-0.16}$ \\
\hline
\end{tabular}
\vspace{-10pt}
\label{tab:LF}
\end{center}
\end{table*}

H$\alpha$ emitters are selected in the following way:
\begin{itemize}
\item The photometric or spectroscopic redshift of the source does not lie in the expected ranges for H$\beta$/[O{\sc iii}]/[O{\sc ii}] emitters ($0.37<z<0.7$ and $0.9<z<1.2$) and 4000\;{\AA} break galaxies ($0.7<z<0.9$) and
\item The photometric or spectroscopic redshift of the source lies in the $0.15<z<0.25$ range.
\end{itemize}
Figure \ref{fig:colcol} displays the colour-colour distribution of line emitters, the cut employed to separate the source types and highlights the location of the H$\alpha$ emitters. All three fields and both filters are shown in the same plot. Separating the data per field and filter results in colour-colour diagrams which are consistent with Figure \ref{fig:colcol}, indicating there are no systematic differences between the populations selected with the two NB filters. The number of H$\alpha$ emitters selected in each field can be found in Table \ref{tab:number}, amounting to a total of $220$ H$\alpha$ emitters. This amounts to almost $40$ per cent of the total number of emitters, as expected and explained in \S\ref{sec:samples:selectionHA}.

\subsubsection{Purity of the H$\alpha$ sample}\label{sec:samples:purity}

We compare the spectroscopic and photometric redshifts in order to study the purity of the H$\alpha$ sample (Figure \ref{fig:zdistrib}). We find that the photometric redshifts are within $0.05$ of the spectroscopic ones. From the sources spectroscopically confirmed to be at lower or higher redshift, none make it into the H$\alpha$ catalogue, implying a very low contamination. Note that the range we used for selecting sources as H$\alpha$ from photometric redshifts is $0.15-0.25$, which is large enough to capture H$\alpha$ emitters in both filters, while minimising contamination. Out of $12$ spectroscopically confirmed emitters we miss two sources, implying completeness higher than $80$ per cent. However, the spectroscopy is limited and the low number statistics could lead to an overestimation or underestimation of the completeness and contamination. Future spectroscopic observations will allow us to further investigate this.

\section{H$\alpha$ luminosity function and star-formation rate density}\label{sec:LHA}

We use the sample of $220$ H$\alpha$ sources to build luminosity functions.

Our filters are sensitive not only to H$\alpha$, but also to the adjacent [N{\sc ii}] double ($6450$ and $6585$ {\AA}) forbidden line. We subtract the [N{\sc ii}] contribution from the line fluxes using the method from \citet{2012MNRAS.420.1926S} to obtain H$\alpha$ fluxes ($F_\mathrm{H{\alpha}}$), which has been spectroscopically confirmed by \citet{2015arXiv150206602S}. The average [N{\sc ii}] contribution is about $30$ per cent of the total line flux.

After we obtain pure H$\alpha$ fluxes $F_\mathrm{H\alpha}$, we calculate the H$\alpha$ luminosity $L_{\mathrm{H}\alpha}$: 
\begin{equation}
\label{eq:L}
L_\mathrm{H\alpha}=4 \pi d^2_{L}(z) F_\mathrm{H\alpha},
\end{equation}
where $d_{L}(z)$ is the luminosity distance ($940$ Mpc for the NB1 filter and $1110$ Mpc for NB2). 

\begin{figure}
\centering
\includegraphics[trim=0cm 0cm 0cm 0cm, width=0.495\textwidth]{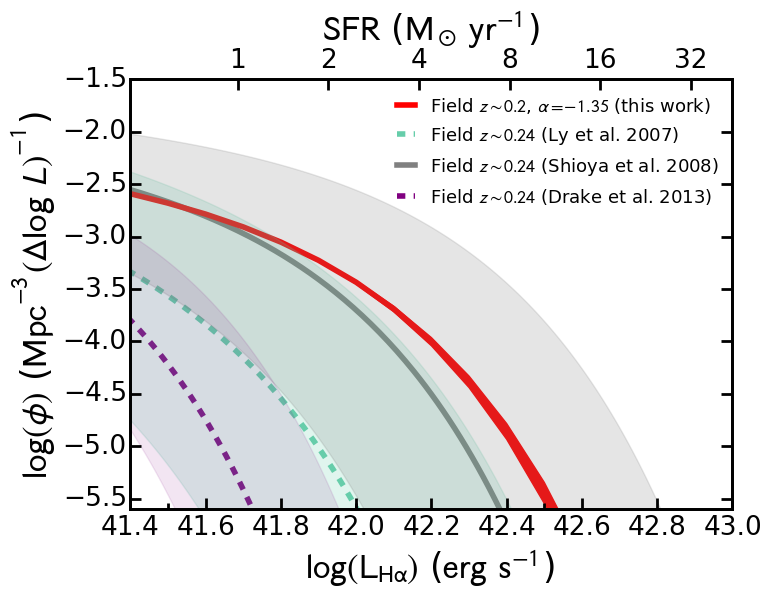}
\caption{A range of luminosity functions at $z\sim0.2$, from the current work and the works of \citet{2008ApJS..175..128S}, \citet{2007ApJ...657..738L} and  \citet{2013MNRAS.433..796D}. In shaded areas, we overplot the ranges allowed by the $1\sigma$ error bars of the LF parameters. The works of  \citet{2008ApJS..175..128S}, \citet{2007ApJ...657..738L} and \citet{2013MNRAS.433..796D} explore the faint end part of the luminosity. The shaded areas indicate the $1\sigma$ uncertainties of the Schechter function parameters. Our measurements are consistent with previous work, but significantly improve the previously unexplored bright end. While our measurement error is given by cosmic variance, as shown in \S\ref{sec:cosmicvariance}. However the other measurement do not include the error given by cosmic variance, which would add an error of about $100-200$ per cent in the parameters.}
\label{fig:LFs02}
\end{figure}

\subsection{Completeness, volume and filter profile corrections}\label{sec:LHA:completeness}
We use the method of \citet{2012MNRAS.420.1926S} to correct for the incompleteness arising from missing sources with faint H$\alpha$ fluxes and/or low $EW$. We select random samples of sources passing the selection criteria for being located at the redshifts traced by the two filters, but which are not selected as H$\alpha$ emitters. Fake H$\alpha$ emission lines are added to these sources which are then passed through the H$\alpha$ selection criteria ($EW$ and $\Sigma$) described at the end of \S\ref{sec:samples:selectionHA}.

Because of the different depth between the pointings and between the four CCD chips, we independently study the recovery rate as function of the H$\alpha$ flux for each chip, pointing, filter and field. The results of the completeness study can be found in the Appendix in Figure \ref{fig:completeness}. Our results are corrected for the effects of incompleteness, especially the H$\alpha$ luminosity function (see \S\ref{sec:HALF}, \S\ref{sec:SFRD} and, e.g., Figures~\ref{fig:bestLF}, \ref{fig:LFs02} and \ref{fig:sfrd}).

The volumes probed in each field and at each redshift assuming that the filters have a perfect top-hat shape are listed in Table \ref{tab:nbobs}. The total co-moving volume probed is $3.63\times10^5$ Mpc$^3$, by far the largest volume ever surveyed in H$\alpha$ at $z\sim0.2$. However, since the filter transmission does not follow perfectly an idealised top hat, we follow the method of \citet{2009MNRAS.398...75S} and \citet{2012MNRAS.420.1926S} and correct the volumes to account for sources missed at the edges of the filter. 

\subsection{Survey limits}\label{sec:LHA:limits}
A $50$ per cent completeness (see Figure~\ref{fig:completeness}) translates to average limiting H$\alpha$ luminosities of $10^{41.1-41.6}$ erg s$^{-1}$ for our survey. This is equivalent to limiting star formation rates (SFR) of $0.5-1.8$ $M_{\odot}$ yr$^{-1}$, with no intrinsic dust extinction applied. If 1 magnitude of dust extinction is applied this is equivalent to $0.2-0.8$ SFR$^*$ (see equation \ref{eq:SFR} in \S\ref{sec:SFRD}). 

The maximum observed H$\alpha$ luminosity our survey probes is $\sim10^{42.4}$ erg s$^{-1}$, equivalent to SFRs of $11$ $M_{\odot}$ yr$^{-1}$ (or $\gtrsim27$ $M_{\odot}$ yr$^{-1}$ if $1$ mag of dust extinction is applied). By comparison, the widest H$\alpha$ survey at a similar redshift, performed by \citet{2008ApJS..175..128S}, reaches $\sim10^{41.9}$ erg s$^{-1}$, or $3.5$ $M_{\odot}$ yr$^{-1}$ ($8.7$  $M_{\odot}$ yr$^{-1}$ with dust extinction). This means our survey probes galaxies more than three times more star forming than previous surveys.

\subsection{H$\alpha$ luminosity function}\label{sec:HALF}

Using our final sample of H$\alpha$ emitters, we build luminosity functions (LF) which characterise the density of sources at any given H$\alpha$ luminosity. To do so, we bin sources based on their luminosity (corrected for the [N{\sc II}] contribution, \S\ref{sec:LHA}, but not for intrinsic dust extinction), by adding their associated inverse co-moving volume, corrected for the real filter profile and incompleteness (as shown in \S\ref{sec:LHA:completeness}). 

We fit the binned data with a \citet{1976ApJ...203..297S} parametrisation:
\begin{equation}
\label{eq:schechter}
\phi(L_{\mathrm{H}\alpha}) \mathrm{d} L_{\mathrm{H}\alpha} = \phi^*\left(\frac{L_{\mathrm{H}\alpha}}{L^*_{\mathrm{H}\alpha}}\right)^{\alpha} e^{-\frac{L_{\mathrm{H}\alpha}}{L^*_{\mathrm{H}\alpha}}} \mathrm{d} \left(\frac{L_{\mathrm{H}\alpha}}{L^*_{\mathrm{H}\alpha}}\right),
\end{equation}
where $L^*_{\mathrm{H}\alpha}$ is the characteristic H$\alpha$ luminosity, $\phi^*$ is the characteristic density of H$\alpha$ emitters and $\alpha$ is the faint-end slope of the LF. Since our data is not deep enough to properly constrain the faint end slope of the LF (see Table \ref{tab:number}), we fix $\alpha$ to two values previously derived in the literature using deep data: $-1.35$ from \citet{2008ApJS..175..128S} and $-1.7$ from \citet{2007ApJ...657..738L}. In fitting the LFs, we assume Poissonian errors.

Our best fit H$\alpha$ LF is described by a typical luminosity $\log(L^*_\mathrm{H\alpha})=10^{(41.71\pm0.02)}$ erg s$^{-1}$ and a characteristic density $\log(\phi^*)=10^{(-2.85\pm0.03)}$ Mpc$^{-3}$ (see Table~\ref{tab:LF} and Figure~\ref{fig:bestLF}). Our data samples really well the bright-end of the LF, which enables us to place tight constraints on $\phi^*$ and $L^*_\mathrm{H\alpha}$ (errors lower than $15$ per cent). However, we lack depth (lowest bin at $\sim 10^{41.4}$ erg s$^{-1}$), so we fix the faint-end slope to $-1.35$, as obtained by \citet{2008ApJS..175..128S} from the previously widest H$\alpha$ survey, which benefits from high-quality, deep data reaching luminosities of $10^{39.3}$ erg s$^{-1}$, but is limited at the bright end. Therefore, the two surveys are highly complementary. Within the overlapping regions with data from both the \citet{2008ApJS..175..128S} and our survey, the measurements are in excellent agreement. However, our LF, constrained up to $L_\mathrm{H\alpha} = 10^{42.5}$ erg s$^{-1}$, indicates a slightly larger value of $L^*_\mathrm{H\alpha}$, but still consistent with \citet{2008ApJS..175..128S} within their large error bars (see Figure \ref{fig:LFs02}). Note that their uncertainties do not include the error from cosmic variance, which can results in $100-200$ errors in the parameters of the LF (see \S\ref{sec:cosmicvariance}). Any discrepancy between the results can be explained by cosmic variance, given Shioya's volume is $\sim10$ times smaller than ours and probes a single field. The differences between the $\phi^*$ results could also be explained by the different colour-colour methods used to separate the H$\alpha$ emitters from higher redshift line emitters.

The discrepancy with other studies is much larger however (see Figure~\ref{fig:LFs02}). Compared to our results, \citet{2007ApJ...657..738L}, slightly overestimate $\phi^*$ (not significant) and underestimate $L^*_\mathrm{H\alpha}$ (at the $2\sigma$ level). \citet{2013MNRAS.433..796D} obtain an $L^*_\mathrm{H\alpha}$ which is highly underestimated ($10^{40.83}$ erg s$^{-1}$). The difference to our value is significant at the $11\sigma$ level. This is entirely driven by Drake's small volume ($\sim30$ times smaller than ours) and the long exposures they were using in their study which prevented the study of sources brighter than $20$ mag in the NB filter. Given the large variations in the LF parameters from cosmic variance, we expect all theses results to be consistent with our measurement, once the cosmic variance error is folded in (see \S\ref{sec:cosmicvariance}).

\subsection{Star formation rate density}\label{sec:SFRD}
We can calculate the star formation rate density (SFRD) at $z\sim0.2$ by integrating the luminosity function and converting H$\alpha$ luminosity to SFR. We use the $L_{\mathrm{H}\alpha}$ to SFR conversion from \citet{1998ARA&A..36..189K}, corrected for the \citet{2003PASP..115..763C} IMF:
\begin{equation}
\label{eq:SFR}
SFR(\mathrm{M_\odot} \mathrm{yr^{-1}}) = 4.4 \times 10^{-42} L_{H\alpha} \mathrm{(erg\, s^{-1})}.
\end{equation}
The luminosity density is obtained by integrating the H$\alpha$ LF:
\begin{align}
\rho_{L_{\mathrm{H}\alpha}} & =   \int_0^{\infty} \phi(L_{\mathrm{H}\alpha})L_{\mathrm{H}\alpha} \mathrm{d}L_{\mathrm{H}\alpha} \\
& =   \Gamma(\alpha+2) \phi^* L^*_{\mathrm{H}\alpha},
\end{align}
where $\Gamma(n)=(n-1)!$ is the Gamma function. By converting from luminosity to SFR through equation \ref{eq:SFR}, the SFRD $\rho_\mathrm{SFR}$ is: 
\begin{equation}
\label{eq:sfrd}
\rho_\mathrm{SFR} =  \Gamma(\alpha+2) \phi^* L^*_\mathrm{H\alpha} 10^{0.4 A_{\mathrm{H}\alpha}} (1-f_\mathrm{AGN})
\end{equation}
where $A_{\mathrm{H}\alpha}$ is the intrinsic H$\alpha$ dust extinction which we assume to be $1$ mag and $f_\mathrm{AGN}=0.15$ is the fraction of the H$\alpha$ luminosity expected to be due to contributions from broad line and narrow line AGN emission \citep[e.g.][]{2010MNRAS.409..421G, 2015MNRAS.450..630S}.

\begin{figure}
\centering
\includegraphics[trim=0cm 0cm 0cm 0cm, width=0.495\textwidth]{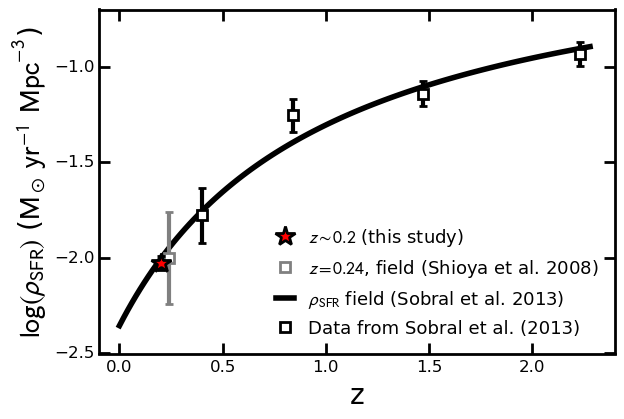}
\caption{Evolution of the SFRD from $z\sim2.23$ to $z\sim0.2$. Our measurement at $z\sim0.2$ confirms the previously discovered decline in SFRD, which can be simply parametrised as a function of redshift \citep{2013MNRAS.428.1128S}.}
\label{fig:sfrd}
\end{figure}

Our measurement of the SFRD, $\rho_\mathrm{SFRD} = 0.0094\pm0.0008$ $\mathrm{M}_\odot \, \mathrm{yr}^{-1} \, \mathrm{Mpc}^{-3}$, which matches with the value of \citet{2008ApJS..175..128S} ($0.010\pm0.006$ $\mathrm{M}_\odot \, \mathrm{yr}^{-1} \, \mathrm{Mpc}^{-3}$). \citet{2013MNRAS.428.1128S} derive a redshift-dependent parametrisation of the SFRD ($\rho_\mathrm{SFRD} = -2.1/(1+z)+\log10(4.4/7.9)$, corrected for the Chabrier IMF) based on their measurements and results from \citet{2007ApJ...657..738L} at $z\sim0.08$ and \citet{2008ApJS..175..128S} at $z\sim0.24$ (see Figure \ref{fig:sfrd}). Our measurement perfectly agrees with the parametrisation, which predicts a value of $0.01$ at $z\sim0.2$. 

\begin{figure*}
\centering
\includegraphics[trim=0cm 0cm 0cm 0cm, width=0.75\textwidth]{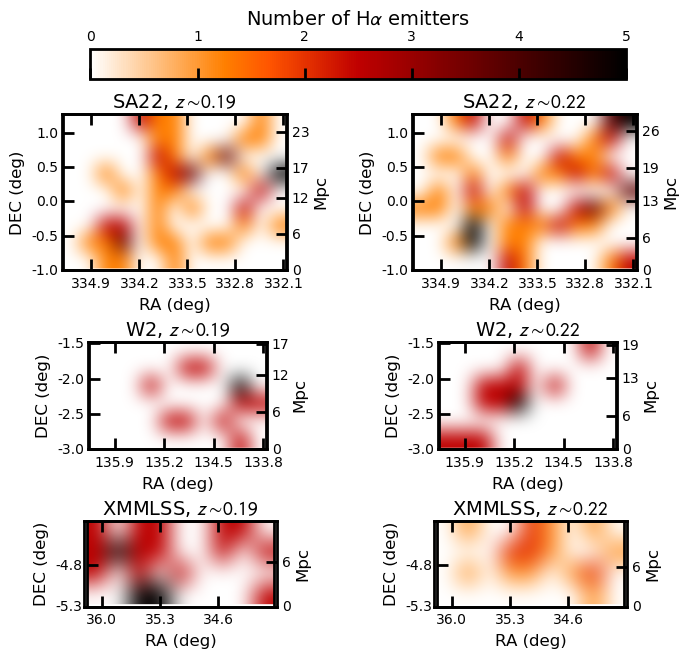}
\caption{Smoothed sky distribution of the H$\alpha$ emitters. Note the amount of cosmic variance within the fields. On average $2$ emitters are found per deg$^2$, but the values vary between $0$ and $5$ sources per deg$^2$.}
\label{fig:sky}
\end{figure*}

\begin{figure}
\centering
\includegraphics[trim=0cm 0cm 0cm 0cm, width=0.495\textwidth ]{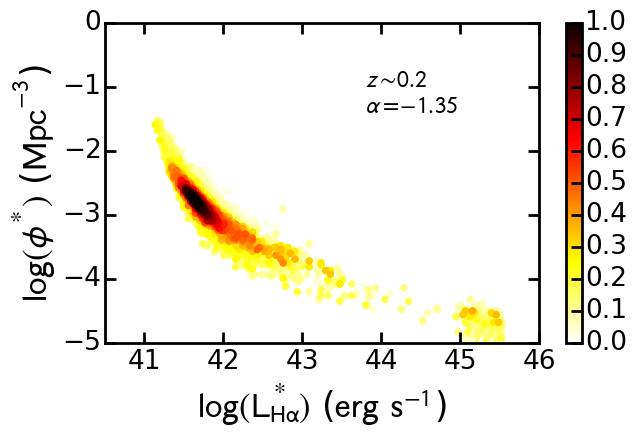}
\caption{The values of LF Schechter parameters $\phi^*$ and $L^*_\mathrm{H\alpha}$, when we fix $\alpha = -1.35$. For fitting the LF, we create $1000$ random sub-samples of H$\alpha$ emitters, at a range of probed volumes. The data points are colour coded with the co-moving volume probed in units of $1000$ Mpc$^3$. Note how at small volumes the scatter is the value is extremely large (up to $4-5$ dex), while at large volumes the values for $\phi^*$ and $L^*_\mathrm{H\alpha}$ converge. We obtain similar results with a different value of $\alpha$ or when we use the data for the two filters separately (see Figure~\ref{fig:LFall})}
\label{fig:LF}
\end{figure}

\begin{figure}
\centering
\includegraphics[trim=0cm 0cm 0cm 0cm, width=0.495\textwidth]{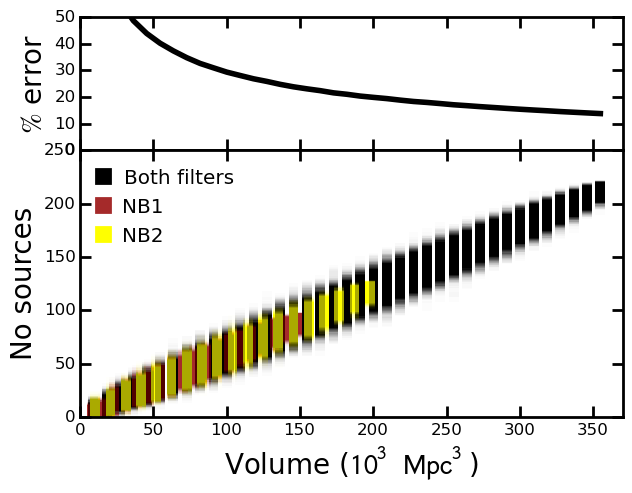}
\caption{Distribution of the number of H$\alpha$ emitters randomly selected within a range of volumes. As expected the larger the volume, the larger the number of sources, with a spread at each volume size caused by cosmic variance. The Poissonian error relative to the mean number of sources does not dominate over the spread caused by cosmic variance, except  where cosmic variance is minimised through the sampling of a large volumes.}
\label{fig:nosources}
\end{figure}

\begin{figure*}
\centering
\begin{subfigure}[b]{0.99\textwidth}
\centering
\includegraphics[trim=0cm 0cm 0cm 0cm, width=0.795\textwidth ]{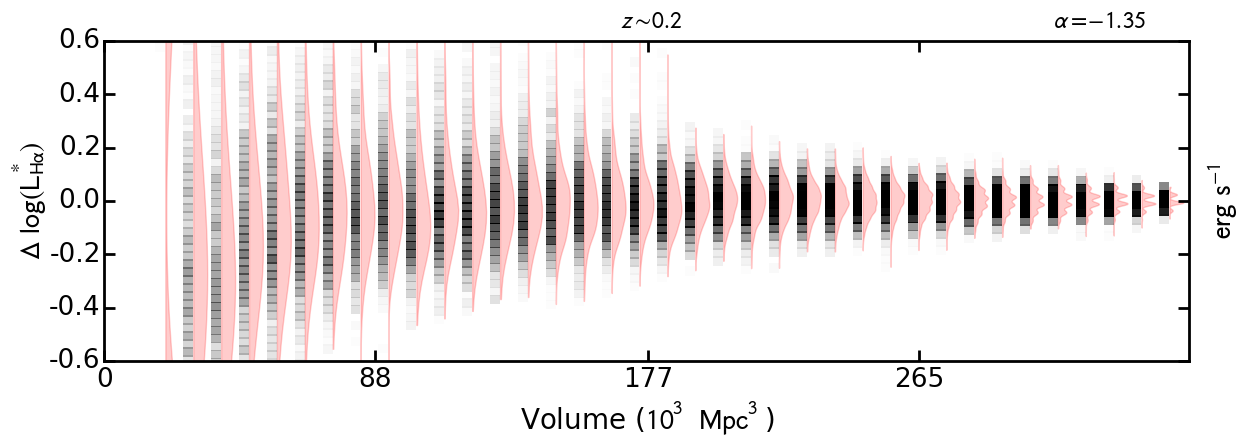}
\end{subfigure}
\begin{subfigure}[b]{0.99\textwidth}
\centering
\includegraphics[trim=0cm 0cm 0cm 0cm, width=0.795\textwidth ]{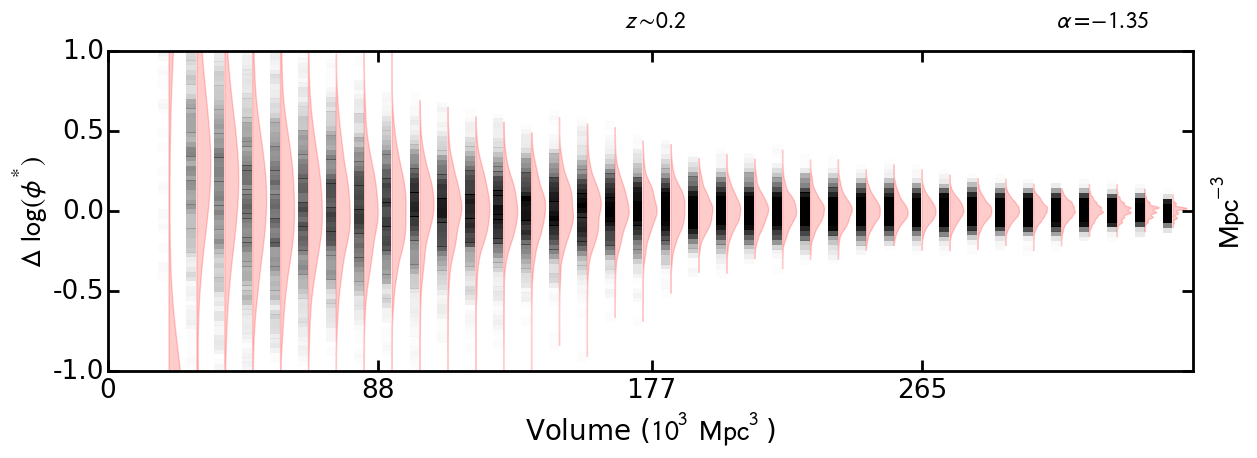}
\end{subfigure}
\caption{The error distribution of the characteristic H$\alpha$ luminosity $L^*_\mathrm{H\alpha}$ and number density $\phi^*$, as function of the volume probed. The error is calculated as fitted value minus the mean of the distribution at each volume. The results are obtained when combining data from both NB filters, with faint end slope fixed to $-1.35$ (see Figures \ref{fig:LL}, \ref{fig:phiphi} and \ref{fig:Lphiall} for results for other $\alpha$ and for the two filters independently). At each volume, $1000$ realisations are performed, based on random samples of sources. Each figure shows the values obtained from the LF fitting in gray-black stripe. Darker colours mean more of the realisations found that particular $L^*_\mathrm{H\alpha}$ or $\phi^*$ value. The violin plot next to each stripe encodes the $L^*_\mathrm{H\alpha}$/$\phi^*$ histogram. The top panel shows the standard deviation $\sigma$ of the $L^*_\mathrm{H\alpha}$ values at each volume size. Note that spread of values drops the larger the volume probed, indicating a convergence in the values of $L^*_\mathrm{H\alpha}$ and $\phi^*$.}
\label{fig:Lphi}
\end{figure*}

\subsection{Distribution of H$\alpha$ emitters}\label{sec:distribution}

Figure \ref{fig:sky} shows the distribution of the H$\alpha$ emitters in the three fields at the two redshifts, as selected in \S\ref{sec:samples:selectionHA}. Note the high degree of cosmic variance within and between the field and at the adjacent redshifts. 

On average, down to a limiting H$\alpha$ luminosity of $10^{41.4}$ erg~$s^{-1}$ or $SFR \sim 1$ $M_\odot$ yr$^{-1}$, we find $\sim2$ H$\alpha$ emitters per square degree (or $\sim3$ per Mpc$^3$). However, there are large areas with no emitters, while parts of the W2 and XMMLSS fields have densities of up to $20$ sources per square degree. The `Sausage' massive, young post-merger galaxy cluster \citet{2014MNRAS.438.1377S, Stroe2015}, where H$\alpha$ emitters were selected with the NB1 filter, was found to be extremely dense in star-forming galaxies and AGN, compared to blank fields. Down to the faintest H$\alpha$ luminosities as our current data surveys ($10^{41.1}$ erg s$^{-1}$), the density is $\sim140$ emitters per square degree, about $70$ times above the average we find over an area of $20$ deg$^2$. Assuming Poissonian noise, the `Sausage' cluster overdensity is significant at the $>11\sigma$ level.

The older `Toothbrush' galaxy cluster merger, where the two subclusters collided about $2$ Gyr ago, behaves differently. The density is about $\sim16$ emitters per square degree, densities similar to the densest parts of our wide, shallow H$\alpha$ survey. Our results thus corroborate the conclusions from \citet{2014MNRAS.438.1377S} and \citet{Stroe2015}.

\begin{table}
\begin{center}
\caption{Bin width $\Delta \log L_\mathrm{H\alpha}$, starting bin $\log L_\mathrm{H\alpha,min}$ and number of bins ($N$) chosen for studying the luminosity function, depending on the volume $V$ probed.}
\vspace{-5pt}
\begin{tabular}{l c c c }
\hline\hline
$V$ range & $\Delta \log L_\mathrm{H\alpha}$  & $\log L_\mathrm{H\alpha,min}$ & $N_\mathrm{bins}$ \\
\hline
$< 2 \times 10^4$ Mpc$^3$ & $0.3$ & $41.5$ & $4$   \\
$2 \times 10^4 - 9 \times 10^5$ Mpc$^3$ & $0.2$ & $41.5$ & $4$ \\
$9 \times 10^5 -18 \times 10^5$ Mpc$^3$ & $0.15$ & $41.4$ & $5$ \\
$18 \times 10^5 - 27 \times 10^4$ Mpc$^3$ & $0.15$ & $41.4$ & $6$ \\
$> 27 \times 10^5$ Mpc$^3$ & $0.1$ & $41.4$ & $8$ \\
\hline
\end{tabular}
\vspace{-10pt}
\label{tab:binning}
\end{center}
\end{table}

\subsection{Quantifying cosmic variance}\label{sec:cosmicvariance}

One of our goals is to understand the impact of cosmic variance and low number statistics on the determination of the LF parameters, especially motivated by the differences in LF found with the previous studies of \citet{2007ApJ...657..738L} and \citet{2013MNRAS.433..796D}. We generate random subsamples of H$\alpha$ emitters, probing a range of volumes. We perform $1000$ realisations starting from the smallest volumes for which we can fit a LF, up to the entire volume of our survey. We perform this experiment using H$\alpha$ emitters in each NB filter and also combine all the data together, following \citet{2015arXiv150206602S}. 

The number of sources for each realisation is plotted in Figure~\ref{fig:nosources}. As expected the average number of sources increases with the volume surveyed. We calculate the standard deviation of the spread in number of sources at each volume and compare that to the Poissonian error. In the calculation of the Poissonian error we take into account the fact that the sources are divided into bins. At very low volumes, the relative Poissonian error dominates over the spread in the number of sources, which is caused by cosmic variance. Given the depth of our survey, at the very small volumes ($<2\times10^4$ Mpc$^3$) the Poissonian error essentially goes to infinity. Overall, the total relative error, calculated as the sum in quadrature of the Poissonian and cosmic variance error, goes down with increasing volume.

Naturally, when surveying a smaller volume, the number of H$\alpha$ sources is proportionally smaller. We therefore adapt the number of bins ($N$), the bin width $\Delta \log L_\mathrm{H\alpha}$ and the starting bin $\log L_\mathrm{H\alpha}$, depending on the volume $V$ probed, as detailed in Table~\ref{tab:binning}.

The results from the different realisations of the LF calculated from H$\alpha$ emitters extracted over a range of volumes can be found in Figure \ref{fig:LF}. At small volumes ($<4\times10^4$ Mpc$^3$), the random realisations of the LF give wildly different results, with values spanning $4-5$ dex. This is driven by two main factors: low number statistics and cosmic variance. The low number of H$\alpha$ emitters in small volumes imposes wide and few $L_\mathrm{H\alpha}$ bins to gain enough number statistics. With few bins, the LF function is barely constrained. Additionally, small volumes do not fully sample the LF at the brightest $L_\mathrm{H\alpha}$, where H$\alpha$ emitters are rare. Therefore, when the volumes are small cosmic variance is significant. However, with the increase of the probed volume, we can much better constrain $\phi^*$ and $L^*_{\mathrm{H}\alpha}$ parameters, by overcoming both Poissonian errors and cosmic variance. This is exemplified in Figure \ref{fig:Lphi}. The standard deviation of the $L^*_{\mathrm{H}\alpha}$ and $\phi^*$ parameters at each volume size becomes smaller with increasing volume. Note however the values of $L^*_{\mathrm{H}\alpha}$ and $\phi^*$ are highly correlated (Figure \ref{fig:LFall}).

As shown in this section, cosmic variance can fully explain the differences found in the literature regarding the H$\alpha$ LF at $z\sim0.2$. By accounting for cosmic variance our LF results can be reconciled with those of \citet{2013MNRAS.433..796D} and \citet{2007ApJ...657..738L}. Our results indicate that at $z\sim0.2$, volumes of at least $10^5$ Mpc$^3$ are required to overcome cosmic variance.

\begin{figure}
\centering
\includegraphics[trim=0cm 0cm 0cm 0cm, width=0.495\textwidth]{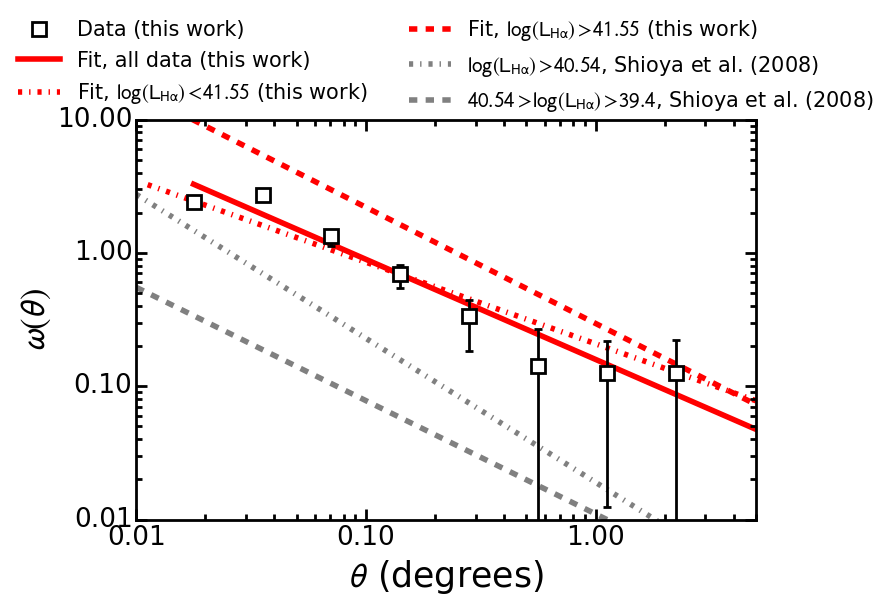}
\caption{Angular two-point correlation function for bright H$\alpha$ emitters ($\mathrm{L_{H\alpha}} \gtrsim 10^{41.0}$ erg s$^{-1}$) at $z\sim0.2$. The best fit power law relation is: $\omega(\theta) = (0.109\pm0.005)\theta^{(-0.79\pm0.04)}$. For comparison, we plot the results for fainter emitters ($\mathrm{L_{H\alpha}} \lesssim 10^{41.5}$ erg s$^{-1}$) from \citet{2008ApJS..175..128S}. We find that more luminous H$\alpha$ emitters are more clustered.}
\label{fig:clustering}
\end{figure}

\begin{table}
\begin{center}
\caption{Two-point correlation function for H$\alpha$ emitters at $z\sim0.2$. Best fit as a single power law of the form $\omega(\theta) = A \theta^\beta$. Note taht the filters and redshift distribution is different for \citet{2008ApJS..175..128S} than for our study, so the amplitudes cannot be directly compared.}
\vspace{-5pt}
\begin{tabular}{l c c c}
\hline\hline
Source & $\log (L_\mathrm{H\alpha})$ (erg s$^{-1}$) & $A$ & $\beta$ \\
\hline
This study &  $41.00-42.40$ & $0.159\pm0.012$ & $-0.75\pm0.05$ \\
Faint & $41.00-41.55$ & $0.208\pm0.035$ & $-0.61\pm0.07$ \\
Bright & $41.55-42.40$ & $0.295\pm0.026$ & $-0.87\pm0.06$ \\
\citet{2008ApJS..175..128S} & $40.54-41.50$ & $0.019\pm0.004$ & $-1.08\pm0.05$\\
\citet{2008ApJS..175..128S} & $39.40-40.54$ & $0.011\pm0.002$ & $-0.85\pm0.05$ \\
\hline
\end{tabular}
\vspace{-10pt}
\label{tab:clustering}
\end{center}
\end{table}

\section{Clustering of H$\alpha$ emitters}\label{sec:clustering}
To study the clustering of our sample of $220$ bright H$\alpha$ emitters at $z\sim0.2$, we start by generating a random catalogue with $1$ million sources. The random catalogue sources follow the geometry of the actual observed fields and masked areas (due to saturated stars) and their number in each CCD of each pointing is normalised according to the depth attained (and hence the density of sources in that area).

We follow the method described in detail in \citet{2010MNRAS.404.1551S}, which evaluates the two-point angular correlation function minimum variance estimator proposed by \citet{1993ApJ...412...64L}:
\begin{equation}
\label{eq:omega}
\omega(\theta) =  1 + \left(\frac{N_R}{N_D} \right)^2 \frac{\mathrm{DD}(\theta)}{\mathrm{RR}(\theta)} - 2 \frac{N_R}{N_D} \frac{\mathrm{DR}(\theta)}{\mathrm{RR}(\theta)},
\end{equation} 
where $\theta$ is the angle on the sky and $N_R$ and $N_D$ are the number of sources in the random and real catalogue of H$\alpha$ sources. $\mathrm{DD}(\theta)$, $\mathrm{RR}(\theta)$ and $\mathrm{DR}(\theta)$ are the number pairs of sources located at distances between $\theta$ and $\theta+\delta\theta$ in the real data, random data and between real and random data, respectively.

Errors on $\omega(\theta)$ are then \citep{1993ApJ...412...64L}:
\begin{equation}
\label{eq:omegaerr}
\Delta\omega(\theta) = \frac{1 + \omega(\theta)}{\sqrt{\mathrm{DD}(\theta)}}.
\end{equation} 

We determine $\omega(\theta)$ using $1000$ different randomly selected sub-samples of sources selected from the randomly generated catalogue. We perform our analysis separately on emitters selected in each filter, but combine the data for the SA22, W2 and XMMLSS fields. We use the full luminosity range ($L_\mathrm{H\alpha} = 10^{41.0-42.4}$ erg s$^{-1}$) of the H$\alpha$ emitters, as well as split the sample in two roughly equal halves: a faint sample with luminosities in the range $10^{41.0-41.55}$ erg s$^{-1}$ and a bright one with luminosities $10^{41.55-42.40}$ erg s$^{-1}$. We bin the data using a range of angular scale bins (with different starting bin $\theta_\mathrm{min}$, bin width $\delta\theta$ and maximum bin $\theta_\mathrm{max}$).

The results are presented in Figure~\ref{fig:clustering} and Table \ref{tab:clustering}. The two-point correlation function for the samples is well described by a single power law. The results for the two filters are considered separately and when combined give fully consistent results within the error bars.

\begin{figure}
\centering
\includegraphics[trim=0cm 0cm 0cm 0cm, width=0.495\textwidth]{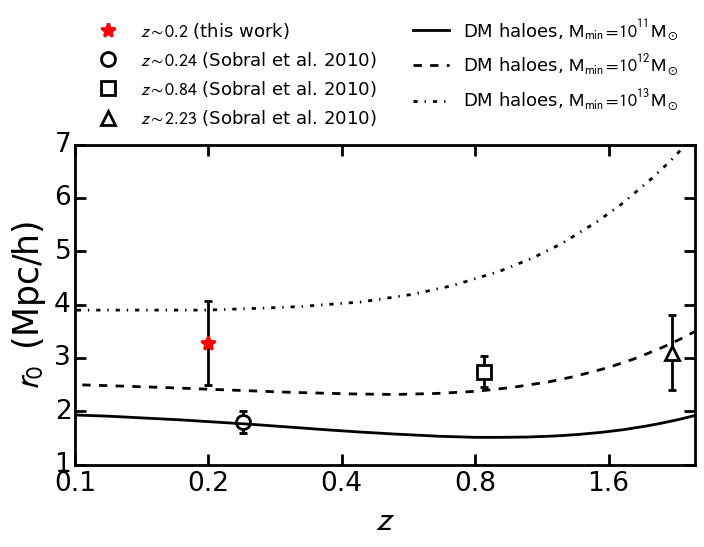}
\caption{The dependence of the clustering length $r_0$ on redshift, using a consistent set of H$\alpha$ emitters selected through NB surveys. For comparison, we are also showing data from \citet{2010MNRAS.404.1551S}. The plot suggests that typical ($L^*_\mathrm{H\alpha}$) emitters have very similar $r_0$ across cosmic time. At $z\sim0.2$, there is a sharp increase in the typical DM halo mass with luminosity of the H$\alpha$ sample. Note however, as shown in Figure \ref{fig:r0ev}, that once corrected for the redshift evolution of the characteristic luminosity, $L_\mathrm{H\alpha}$ sets the position of galaxies in relation to DM halo host. }
\label{fig:r0}
\end{figure}

\begin{figure*}
\centering
\includegraphics[trim=0cm 0cm 0cm 0cm, width=0.495\textwidth]{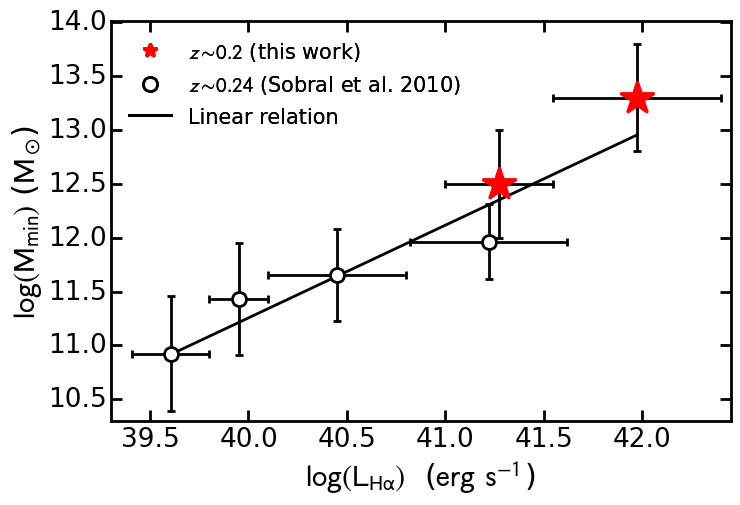}
\includegraphics[trim=0cm 0cm 0cm 0cm, width=0.495\textwidth]{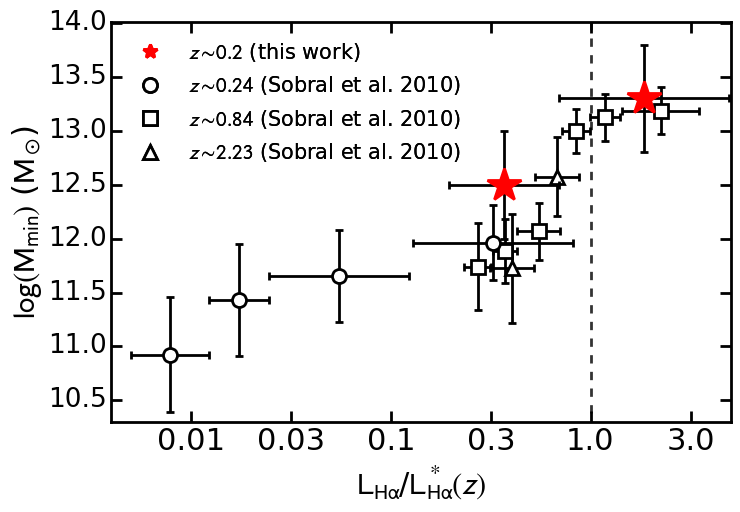}
\caption{The minimum DM halo mass ($M_\mathrm{DM}$) as function of luminosity ($L_\mathrm{H\alpha}$, left) and luminosity scaled by the characteristic luminosity at the respective redshift ($L_\mathrm{H\alpha}/L^*_\mathrm{H\alpha}(z)$, right). The data from \citet{2010MNRAS.404.1551S}, split per luminosity bin, are shown for comparison. The $\sim0.2$ points are renormalised using the $L^*_\mathrm{H\alpha}$ derived in this paper. All luminosities are not corrected for intrinsic dust extinction. Note the relation between the H$\alpha$ luminosity and host mass. When scaled for the typical luminosity, a clear relation between DM halo mass and luminosity is observed from $z\sim2.23$ to $z\sim0.2$.}
\label{fig:r0ev}
\end{figure*}

Note we studied only the range $0.02 \ \mathrm{deg}<\theta<3.0 \ \mathrm{deg}$, where there was enough signal. At scales smaller than $<0.02 \ \mathrm{deg}$, a flattening of $\omega(\theta)$ occurs, maybe caused by bright H$\alpha$ emitters not being able to reside in a single halo. Additionally, since our survey is not very deep, we do not probe the regime where satellites are expected. Therefore, we cannot evaluate the departure of the two-point correlation function from a single power, which is caused by the transition from the large scale (two galaxies residing in separate dark matter halo) to the small scale clustering regime \citep[galaxies sharing a single halo, e.g.][]{2005ApJ...620L...1O}.

Previous research indicates that bright H$\alpha$ galaxies as well as Lyman break galaxies are more clustered than the faint ones. \citet{2008ApJS..175..128S} found that the two-point correlation function for faint H$\alpha$ emitters ($\mathrm{L_{H\alpha}}<10^{40.54}$ erg s$^{-1}$) at $z\sim0.24$ follows the relationship: $\omega(\theta) = (0.011\pm0.002) \theta^{(-0.84\pm0.05)}$, while brighter emitters with $10^{40.54} < \mathrm{L_{H\alpha}} \lesssim 10^{41.5}$ erg s$^{-1}$ follow the relationship: $\omega(\theta) = (0.019\pm0.004) \theta^{(-1.08\pm0.05)}$. The amplitude of the two-point correlation function for our faint sample is $0.208\pm0.035$, while for the bright sample it is slightly larger: $0.295\pm0.026$. The relation is also steeper for the bright sample than for the faint sample. Our results therefore support and extend the claim that brighter (and hence more star-forming galaxies) are more clustered than faint ones to very high luminosities beyond $10^{41.0}$ erg s$^{-1}$ up to $10^{42.4}$ erg s$^{-1}$ ($L/L^*_\mathrm{H\alpha} \sim 5.0$).

We use the inverse Limber transformation and the redshift distribution of the NB filters to translate the two-point correlation function into a 3D spatial correlation \citep{1980lssu.book.....P}, assuming the latter is well described by $\epsilon = (r/r_0)^\gamma$, where $r_0$ is the real-space correlation length of the H$\alpha$ emitters. Following the method of \citet{2010MNRAS.404.1551S}, we assume that the two filters have a perfect top-hat shape. We compute $r_0$ for each realisation of $\omega(\theta)$ in each filter, by fixing $\beta=-0.8$. We finally combine the data for the two filters. The dependence of $r_0$ on redshift is shown in Figure \ref{fig:r0}. 

For the full sample, we obtain a correlation length $r_0 = 3.3$ Mpc/h with a standard deviation $0.8$ Mpc/h. We obtain $r_0=3.5\pm1.1$ Mpc/h for our fainter H$\alpha$ sample and $5.0\pm1.5$ Mpc/h for the brighter one. Our measurements are larger than those of \citet{2010MNRAS.404.1551S} at $z\sim0.24$ (based on the sample from \citet{2008ApJS..175..128S}), which find a value of $1.8\pm0.2$ Mpc/h for their sample with $10^{39.4}< L_\mathrm{H\alpha} < 10^{41.5}$ erg s$^{-1}$. As expected, fainter H$\alpha$ galaxies have smaller correlation lengths than brighter ones \citep{2001MNRAS.328...64N, 2008ApJS..175..128S, 2010MNRAS.404.1551S}. The correlation length also depends on redshift, but the evolution is driven by the typical luminosity: at high redshift, H$\alpha$ emitters are on average brighter and have larger $r_0$ than lower redshift sources. 

Similar results are found by \citet{2010MNRAS.407.1212H}, who select galaxies using $K$ band luminosity as proxy for stellar mass. The authors find that red galaxies, likely mostly ellipticals, are more clustered than the blue galaxies. Selecting star-forming galaxies based on colours, they find that $r_0$ drops with redshift. However, no dependence of $r_0$ on broad band luminosity was found. By contrast, \citet{2014A&A...568A..24B} use a mass selected sample and find that higher mass galaxies tend to have larger clustering lengths. Additionally, they find that the clustering strength increases with stellar mass. Stellar mass correlates well with SFR \citep[e.g. at $z\sim0.2$][]{Stroe2015}, which can then be translated to an equivalent H$\alpha$ luminosity though equation \ref{eq:SFR}. The results from \citet{2014A&A...568A..24B} may indicate that more star forming, more luminous galaxies have larger $r_0$ which is consistent with our findings. Note however that \citet{2010MNRAS.404.1551S} controlled for both H$\alpha$ luminosity and mass ($K$ band luminosity) and found both are important for the evolution of $r_0$: $r_0$ increases with both higher $L_\mathrm{H\alpha}$ and $K$ band luminosity.

The clustering of the H$\alpha$ emitters depends on the clustering of their host dark matter (DM) haloes. The bias parameter $b(z)$ describes how the matter distribution traces the DM distribution, as function of redshift. In the bias model of \citet{1997MNRAS.286..115M}, the physical parameters of galaxies are determined by their host dark matter halo mass. In such a model, $b(z)$ depends on the minimum mass of the DM halo. Figure \ref{fig:r0} also contain $r_0$ predictions for dark matter (DM) haloes with fixed minimum mass of $M_\mathrm{min} = 10^{11-13}M_\odot$, as calculated by \citet{2008MNRAS.388.1473G} assuming a $\Lambda$CDM cosmology and an evolving bias model from \citet{1997MNRAS.286..115M} and \citet{1998MNRAS.299...95M}. Note however, the $r_0$ prediction is highly dependent on the model, see for example \citet{2010MNRAS.407.1212H}. We thus note that while the trends are valid, the normalisation of the $M_\mathrm{min}$ could be higher than that used here, leading to lower masses than derived here.

The emitters from \citet{2008ApJS..175..128S}, probing fainter H$\alpha$ regimes with $L_\mathrm{H\alpha}<10^{41.5}$ erg s$^{-1}$, reside in DM haloes of $10^{11} M_\odot$ mass. These are most likely dwarf galaxies. By contrast, our faint sample is hosted by DM haloes of about $10^{12.5} M_\odot$ mass, about the mass of the Milky Way. The bright H$\alpha$ emitters are hosted by $\sim10^{13-13.5} M_\odot$ DM haloes, which are most probably already galaxy groups. 

Figure \ref{fig:r0ev} shows how the DM halo minimum mass varies as function of H$\alpha$ luminosity and the luminosity scaled by the characteristic luminosity at that redshift ($L_\mathrm{H\alpha}/L^*_\mathrm{H\alpha}(z)$). By comparing our results, with the results from \citet{2010MNRAS.404.1551S} (based on data from \citet{2008ApJS..175..128S}), we find a linear correlation between the host minimum DM halo mass and luminosity (in log-log space, see Figure \ref{fig:r0ev}). This indicates that more luminous, more star-forming galaxies reside is more massive dark matter haloes.

Accounting for the evolution of the characteristic luminosity with redshift, we find that more luminous emitters reside in more massive DM haloes, irrespective of redshift. Such a comparison between $z<0.4$ and $z>4$ samples has been previously difficult because of the different $L_\mathrm{H\alpha}/L^*_\mathrm{H\alpha}(z)$ ranges probed in the different redshift ranges. With our measurements, we probe beyond $L^*_\mathrm{H\alpha}$ at $z\sim0.2$ for the first time to be fully comparable with samples up to $z\sim2.23$. Our measurements therefore confirm the results from \citet{2010MNRAS.404.1551S} and \citet{2012MNRAS.426..679G}, who find that $L^*_\mathrm{H\alpha}$ galaxies reside in $\sim10^{13}$ $M_\odot$, Milky Way size DM haloes, at all redshifts. The results indicate the the position of a star forming galaxies within the H$\alpha$ luminosity function is dictated by the host DM halo mass, at all cosmic times since $\sim2.3$. 

\section{Conclusions}\label{sec:conclusion}
In order to constrain the evolution of the star-forming galaxies across cosmic time, large samples of sources are necessary. Such samples are available at high redshifts ($z>0.8$) through NB selected H$\alpha$ emitter samples which probe large volumes ($>10^5$ Mpc$^3$) and overcome cosmic variance. However, at low redshifts ($z<0.8$), large areas ($>15$ deg$^2$) need to be surveyed in order to match the volumes at high redshift. By carrying out the largest survey of H$\alpha$ emitters at $z\sim0.2$, we produce a luminosity function describing typical galaxies within representative volumes of the Universe. With our large sample of bright emitters we study their distribution and clustering and place it in the context of the evolution of the SFRD throughout cosmic history. Our main results are:
\begin{itemize}
\item The H$\alpha$ luminosity function at $z\sim0.2$ is well described by a Schechter function with $\log(\phi^*)=-2.85\pm0.03$ (Mpc$^{-3}$) and $\log(L^*_\mathrm{H\alpha})=41.71\pm0.02$ (erg s$^{-1}$). We find that previous studies, probing far smaller volumes, underestimate the characteristic luminosity $L^*_\mathrm{H\alpha}$, but are reconciled with our results if cosmic variance uncertainties are taken into account. For volumes typically probed in previous H$\alpha$ works at $z\sim0.2$ of $<5\times10^4$ Mpc$^3$, cosmic variance can account to more than $50$ per cent variance in the LF parameters.
\item By assuming a $15$ per cent AGN fraction, we derive a star formation rate density of $\rho_\mathrm{SFRD}=0.0094\pm0.0008$ M$_\odot$ yr$^{-1}$ Mpc$^{-1}$.
\item We find significant cosmic variance in the distribution of the H$\alpha$ emitters, but on average $1-4$ bright ($L_\mathrm{H\alpha}>10^{41.1}$ erg s$^{-1}$) H$\alpha$ emitters are found per square degree.
\item We study the clustering of the H$\alpha$ emitters. The two-point correlation function is well fit by a single power law $\omega(\theta) = (0.159\pm0.012) \theta^{(-0.75\pm0.05)}$, with a spatial clustering length $r_0 = 5.0\pm1.1$ Mpc/h for the bright sample ($10^{41.0-41.55}$ erg s$^{-1}$) and $r_0 = 3.5\pm1.1$ Mpc/h for the faint sample ($10^{41.55-42.40}$ erg s$^{-1}$). Our results confirm that luminous, strongly star-forming galaxies are more clustered than those weakly star-forming.
\item We find that, at $z\sim0.2$, the higher the SFR, the more massive the DM halo host is. When accounting for the redshift dependence of the characteristic H$\alpha$ luminosity, there is no redshift dependence of the host mass, but a strong dependence on $L_\mathrm{H\alpha}/L^*_\mathrm{H\alpha}(z)$.  
\end{itemize}

\section*{Acknowledgements}
We thank the referee for comments which improved the clarity and interpretation of our results. Based on observations made with the Isaac Newton Telescope (proposal I13BN008) operated on the island of La Palma by the Isaac Newton Group in the Spanish Observatorio del Roque de los Muchachos of the Instituto de Astrof{\'i}sica de Canarias. Based on observations obtained with MegaPrime/MegaCam, a joint project of CFHT and CEA/IRFU, at the Canada-France-Hawaii Telescope (CFHT) which is operated by the National Research Council (NRC) of Canada, the Institut National des Science de l'Univers of the Centre National de la Recherche Scientifique (CNRS) of France, and the University of Hawaii. This work is based in part on data products produced at Terapix available at the Canadian Astronomy Data Centre as part of the Canada-France-Hawaii Telescope Legacy Survey, a collaborative project of NRC and CNRS. Based on observations obtained as part of the VISTA Hemisphere Survey, ESO Progam, 179.A-2010 (PI: McMahon). Based on data products from observations made with ESO Telescopes at the La Silla or Paranal Observatories under ESO programme ID 179.A-2006. The UKIDSS project is defined in \citet{2007MNRAS.379.1599L}. UKIDSS uses the UKIRT Wide Field Camera \citep[WFCAM;][]{2007A&A...467..777C}. The photometric system is described in \citet{2006MNRAS.367..454H}, and the calibration is described in \citet{2009MNRAS.394..675H}. The pipeline processing and science archive are described in Irwin et al (2009, in prep) and Hambly et al (2008). This research has made use of the NASA/IPAC Extragalactic Database (NED) which is operated by the Jet Propulsion Laboratory, California Institute of Technology, under contract with the National Aeronautics and Space Administration. This research has made use of NASA's Astrophysics Data System. AS acknowledges financial support from an NWO top subsidy (614.001.006). DS acknowledges financial support from the Netherlands Organisation for Scientific research (NWO) through a Veni fellowship, from FCT through a FCT Investigator Starting Grant and Start-up Grant (IF/01154/2012/CP0189/CT0010) and from FCT grant PEst-OE/FIS/UI2751/2014. 

\bibliographystyle{mnras.bst}
\bibliography{Halpha_wide}

\begin{thebibliography}{}

\bibitem[Ahn et al.(2012)]{2012ApJS..203...21A}
Ahn, C.~P., Alexandroff, R., Allende Prieto, C., et al.\ 2012, ApJS, 203, 21 

\bibitem[Bertin \& Arnouts(1996)]{1996A&AS..117..393B}
Bertin, E., \& Arnouts, S.\ 1996, A\&AS, 117, 393 

\bibitem[Bertin et al.(2002)]{2002ASPC..281..228B}
Bertin, E., Mellier, Y., Radovich, M., et al.\ 2002, Astronomical Data Analysis Software and Systems XI, 281, 228 

\bibitem[Bertin(2006)]{2006ASPC..351..112B}
Bertin, E.\ 2006, Astronomical Data Analysis Software and Systems XV, 351, 112 

\bibitem[Best et al.(2010)]{2010arXiv1003.5183B}
Best, P., Smail, I., Sobral, D., et al.\ 2010, arXiv:1003.5183 

\bibitem[Bielby et al.(2014)]{2014A&A...568A..24B}
Bielby, R.~M., Gonzalez-Perez, V., McCracken, H.~J., et al.\ 2014, A\&A, 568, A24 

\bibitem[Bland-Hawthorn et al.(2001)]{2001ApJ...563..611B} 
Bland-Hawthorn, J., van Breugel, W., Gillingham, P.~R., Baldry, I.~K., \& Jones, D.~H.\ 2001, ApJ, 563, 611 

\bibitem[Bouwens et al.(2011)]{2011ApJ...737...90B}
Bouwens, R.~J., Illingworth, G.~D., Oesch, P.~A., et al.\ 2011, ApJ, 737, 90 

\bibitem[Bouwens et al.(2015)]{2015ApJ...803...34B}
Bouwens, R.~J., Illingworth, G.~D., Oesch, P.~A., et al.\ 2015, ApJ, 803, 34 

\bibitem[{Bunker} {et~al.}(1995)]{1995MNRAS.273..513B}
{Bunker}, {A.~J.}, {Warren}, {S.~J.}, {Hewett}, {P.~C.}, \& {Clements}, {D.~L.}\ 1995, MNRAS, 273, 513 

\bibitem[Casali et al.(2007)]{2007A&A...467..777C}
Casali, M., Adamson, A., Alves de Oliveira, C., et al.\ 2007, A\&A, 467, 777 

\bibitem[Chabrier(2003)]{2003PASP..115..763C}
Chabrier, G.\ 2003, PASP, 115, 763 

\bibitem[Dale et al.(2010)]{2010ApJ...712L.189D}
Dale, D.~A., Barlow, R.~J., Cohen, S.~A., et al.\ 2010, ApJL, 712, L189 

\bibitem[Drake et al.(2013)]{2013MNRAS.433..796D}
Drake, A.~B., Simpson, C., Collins, C.~A., et al.\ 2013, MNRAS, 433, 796 

\bibitem[Dressler(1980)]{1980ApJ...236..351D}
{Dressler}, A.\ 1980, ApJ, 236, 351 

\bibitem[Drury(1983)]{1983RPPh...46..973D}
{Drury}, {L.~O.} 1983, Reports on Progress in Physics, 46, 973

\bibitem[Erben et al.(2013)]{2013MNRAS.433.2545E}
Erben, T., Hildebrandt, H., Miller, L., et al.\ 2013, MNRAS, 433, 2545 

\bibitem[Garcet et al.(2007)]{2007A&A...474..473G}
Garcet, O., Gandhi, P., Gosset, E., et al.\ 2007, A\&A, 474, 473 

\bibitem[Garn \& Best(2010)]{2010MNRAS.409..421G}
Garn, T., \& Best, P.~N.\ 2010, MNRAS, 409, 421 

\bibitem[Geach et al.(2008)]{2008MNRAS.388.1473G}
Geach, J.~E., Smail, I., Best, P.~N., et al.\ 2008, MNRAS, 388, 1473 

\bibitem[Geach et al.(2012)]{2012MNRAS.426..679G}
Geach, J.~E., Sobral, D., Hickox, R.~C., et al.\ 2012, MNRAS, 426, 679 

\bibitem[Gunawardhana et al.(2013)]{2013MNRAS.433.2764G}
Gunawardhana, M.~L.~P., Hopkins, A.~M., Bland-Hawthorn, J., et al.\ 2013, MNRAS, 433, 
2764 

\bibitem[Gwyn(2012)]{2012AJ....143...38G}
Gwyn, S.~D.~J.\ 2012, AJ, 143, 38 

\bibitem[Hartley et al.(2010)]{2010MNRAS.407.1212H}
Hartley, W.~G., Almaini, O., Cirasuolo, M., et al.\ 2010, MNRAS, 407, 1212 

\bibitem[Hewett et al.(2006)]{2006MNRAS.367..454H}
Hewett, P.~C., Warren, S.~J., Leggett, S.~K., \& Hodgkin, S.~T.\ 2006, MNRAS, 367, 454 

\bibitem[Hodgkin et al.(2009)]{2009MNRAS.394..675H}
Hodgkin, S.~T., Irwin, M.~J., Hewett, P.~C., \& Warren, S.~J.\ 2009, MNRAS, 394, 675 

\bibitem[Ilbert et al.(2006)]{2006A&A...457..841I}
Ilbert, O., Arnouts, S., McCracken, H.~J., et al.\ 2006, A\&A, 457, 841 

\bibitem[Jarvis et al.(2013)]{2013MNRAS.428.1281J}
Jarvis, M.~J., Bonfield, D.~G., Bruce, V.~A., et al.\ 2013, MNRAS, 428, 1281 

\bibitem[Karim et al.(2011)]{2011ApJ...730...61K}
Karim, A., Schinnerer, E., Mart{\'{\i}}nez-Sansigre, A., et al.\ 2011, ApJ, 730, 61 

\bibitem[Kennicutt(1998)]{1998ARA&A..36..189K}
Kennicutt, R.~C., Jr.\ 1998, ARA\&A, 36, 189 

\bibitem[Kirkpatrick et al.(1991)]{1991ApJS...77..417K}
Kirkpatrick, J.~D., Henry, T.~J., \& McCarthy, D.~W., Jr.\ 1991, ApJS, 77, 417 

\bibitem[Kirkpatrick et al.(1999)]{1999ApJ...519..802K}
Kirkpatrick, J.~D., Reid, I.~N., Liebert, J., et al.\ 1999, ApJ, 519, 802 

\bibitem[Kurk et al.(2004)]{2004A&A...428..793K}
Kurk, J.~D., Pentericci, L., R{\"o}ttgering, H.~J.~A., \& Miley, G.~K.\ 2004, A\&A, 428, 793 

\bibitem[Landy \& Szalay(1993)]{1993ApJ...412...64L}
Landy, S.~D., \& Szalay, A.~S.\ 1993, ApJ, 412, 64 

\bibitem[Lawrence et al.(2007)]{2007MNRAS.379.1599L}
Lawrence, A., Warren, S.~J., Almaini, O., et al.\ 2007, MNRASs, 379, 1599 

\bibitem[Lilly et al.(1996)]{1996ApJ...460L...1L}
Lilly, S.~J., Le Fevre, O., Hammer, F., \& Crampton, D.\ 1996, ApJL, 460, L1 

\bibitem[Ly et al.(2007)]{2007ApJ...657..738L}
Ly, C., Malkan, M.~A., Kashikawa, N., et al.\ 2007, ApJ, 657, 738 

\bibitem[Matarrese et al.(1997)]{1997MNRAS.286..115M}
Matarrese, S., Coles, P., Lucchin, F., \& Moscardini, L.\ 1997, MNRAS, 286, 115 

\bibitem[Matthee et al.(2014)]{2014MNRAS.440.2375M}
Matthee, J.~J.~A., Sobral, D., Swinbank, A.~M., et al.\ 2014, MNRAS, 440, 2375 

\bibitem[Melnyk et al.(2013)]{2013A&A...557A..81M}
Melnyk, O., Plionis, M., Elyiv, A., et al.\ 2013, A\&A, 557, AA81 

\bibitem[Moorwood et al.(2000)]{2000A&A...362....9M}
Moorwood, A.~F.~M., van der Werf, P.~P., Cuby, J.~G., \& Oliva, E.\ 2000, A\&A, 362, 9 

\bibitem[Moscardini et al.(1998)]{1998MNRAS.299...95M}
Moscardini, L., Coles, P., Lucchin, F., \& Matarrese, S.\ 1998, MNRAS, 299, 95 

\bibitem[Naylor(1998)]{1998MNRAS.296..339N}
Naylor, T.\ 1998, MNRAS, 296, 339 

\bibitem[Norberg et al.(2001)]{2001MNRAS.328...64N} Norberg, P., Baugh, 
C.~M., Hawkins, E., et al.\ 2001, MNRAS, 328, 64 

\bibitem[Ouchi et al.(2005)]{2005ApJ...620L...1O}
Ouchi, M., Shimasaku, K., Akiyama, M., et al.\ 2005, ApJL, 620, L1 

\bibitem[Peebles(1980)]{1980lssu.book.....P}
Peebles, P.~J.~E.\ 1980, Research supported by the National Science Foundation.~Princeton, N.J., Princeton University Press, 1980.~435 p.

\bibitem[Pierre et al.(2004)]{2004JCAP...09..011P}
Pierre, M., Valtchanov, I., Altieri, B., et al.\ 2004, JCAP, 9, 011 

\bibitem[Polletta et al.(2007)]{2007ApJ...663...81P} 
Polletta, M., Tajer, M., Maraschi, L., et al.\ 2007, ApJ, 663, 81 

\bibitem[Roberts(1962)]{1962AJ.....67..437R}
Roberts, M.~S.\ 1962, AJ, 67, 437 

\bibitem[Schechter(1976)]{1976ApJ...203..297S}
Schechter, P.\ 1976, ApJ, 203, 297 

\bibitem[Shioya et al.(2008)]{2008ApJS..175..128S}
Shioya, Y., Taniguchi, Y., Sasaki, S.~S., et al.\ 2008, ApJS, 175, 128 

\bibitem[{Sobral} {et~al.}(2009)]{2009MNRAS.398...75S}
{Sobral}, {D.}, {Best}, {P.~N.}, {Geach}, {J.~E.}, {et~al.} 2009, MNRAS, 398, 75 

\bibitem[Sobral et al.(2010)]{2010MNRAS.404.1551S}
Sobral, D., Best, P.~N., Geach, J.~E., et al.\ 2010, MNRAS, 404, 1551 

\bibitem[Sobral et al.(2012)]{2012MNRAS.420.1926S}
Sobral, D., Best, P.~N., Matsuda, Y., et al.\ 2012, MNRAS, 420, 1926 

\bibitem[Sobral et al.(2013)]{2013MNRAS.428.1128S}
Sobral, D., Smail, I., Best, P.~N., et al.\ 2013, MNRAS, 428, 1128 

\bibitem[Sobral et al.(2014)]{2014MNRAS.437.3516S}
Sobral, D., Best, P.~N., Smail, I., et al.\ 2014, MNRAS, 437, 3516 

\bibitem[Sobral et al.(2015a)]{2015MNRAS.450..630S}
Sobral, D., Stroe, A., Dawson, W.~A., et al.\ 2015a, MNRAS, 450, 630 

\bibitem[Sobral et al.(2015b)]{2015arXiv150206602S}
Sobral, D., Matthee, J., Best, P.~N., et al.\ 2015b, arXiv:1502.06602 

\bibitem[Stott et al.(2013)]{2013MNRAS.430.1158S}
Stott, J.~P., Sobral, D., Smail, I., et al.\ 2013, MNRAS, 430, 1158 

\bibitem[Stroe et al.(2014)]{2014MNRAS.438.1377S}
Stroe, A., Sobral, D., R{\"o}ttgering, H.~J.~A., \& van Weeren, R.~J.\ 2014, MNRAS, 438, 1377 

\bibitem[Stroe et al.(2015)]{Stroe2015}
Stroe, A., Sobral, D., Dawson, W., et al.\ 2015, MNRAS, 450, 646 

\bibitem[Tajer et al.(2007)]{2007A&A...467...73T}
Tajer, M., Polletta, M., Chiappetti, L., et al.\ 2007, A\&A, 467, 73 

\bibitem[Warren et al.(2007)]{2007astro.ph..3037W}
Warren, S.~J., Cross, N.~J.~G., Dye, S., et al.\ 2007, arXiv:astro-ph/0703037 

\bibitem[Zacharias et al.(2013)]{2013AJ....145...44Z}
Zacharias, N., Finch, C.~T., Girard, T.~M., et al.\ 2013, AJ, 145, 44 

\end{thebibliography}

\appendix
\section{Survey completeness}
\begin{figure*}
\centering
\begin{subfigure}[b]{0.32\textwidth}
\includegraphics[trim=0cm 0cm 0cm 0cm, width=0.995\textwidth]{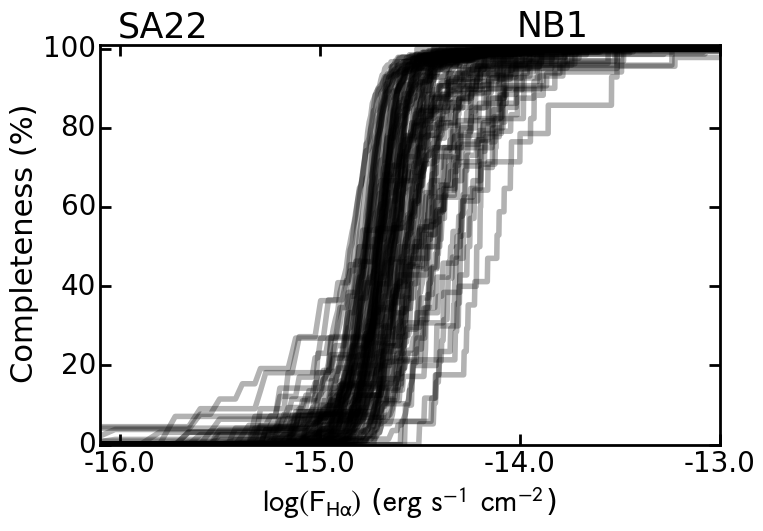}
\end{subfigure}
\hspace{5pt}
\begin{subfigure}[b]{0.32\textwidth}
\includegraphics[trim=0cm 0cm 0cm 0cm, width=0.995\textwidth ]{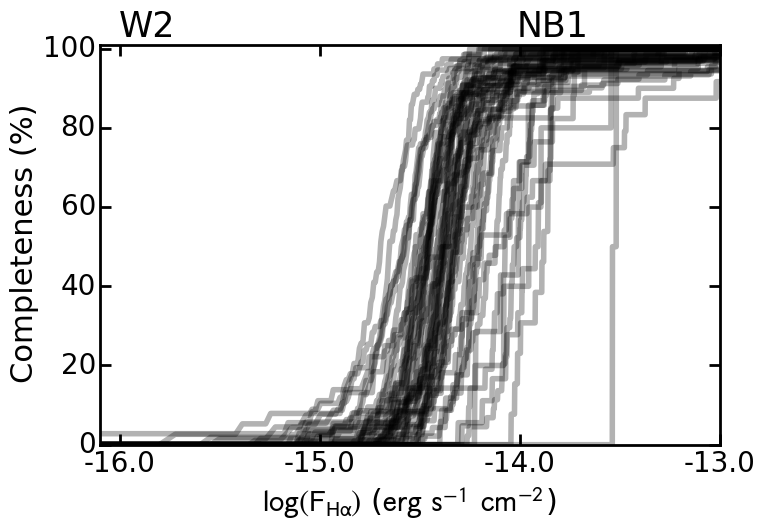}
\end{subfigure}
\begin{subfigure}[b]{0.32\textwidth}
\includegraphics[trim=0cm 0cm 0cm 0cm, width=0.995\textwidth ]{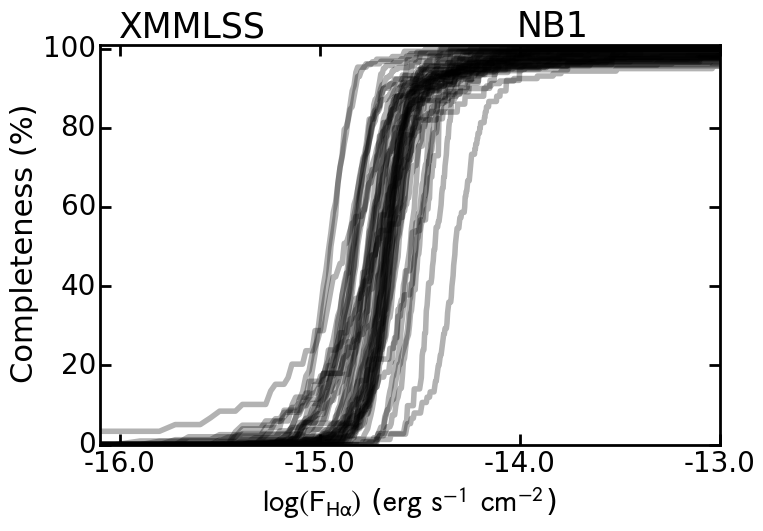}
\end{subfigure}
\hspace{5pt}
\begin{subfigure}[b]{0.32\textwidth}
\includegraphics[trim=0cm 0cm 0cm 0cm, width=0.995\textwidth ]{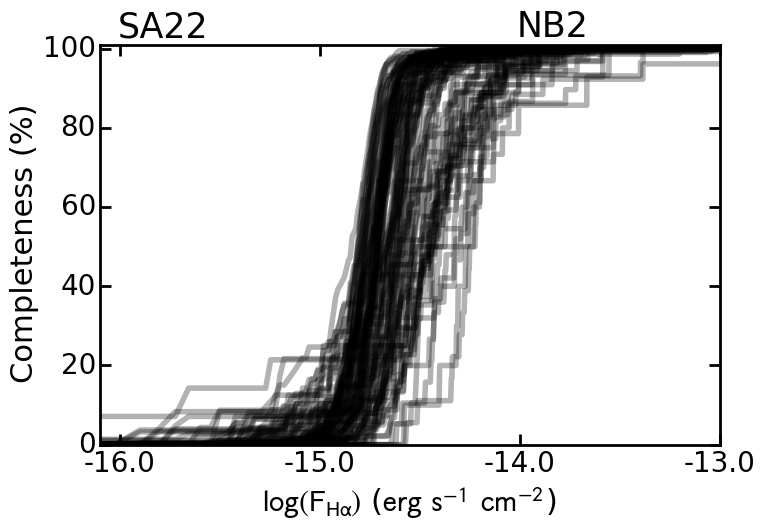}
\end{subfigure}
\begin{subfigure}[b]{0.32\textwidth}
\includegraphics[trim=0cm 0cm 0cm 0cm, width=0.995\textwidth ]{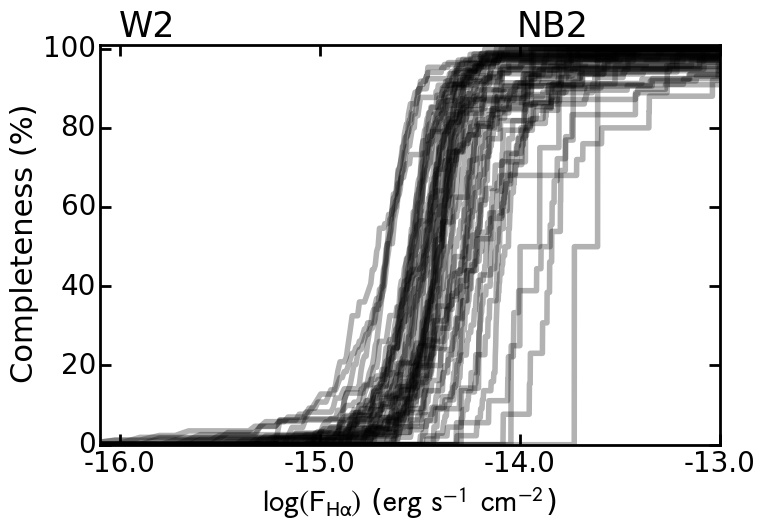}
\end{subfigure}
\hspace{5pt}
\begin{subfigure}[b]{0.32\textwidth}
\includegraphics[trim=0cm 0cm 0cm 0cm, width=0.995\textwidth ]{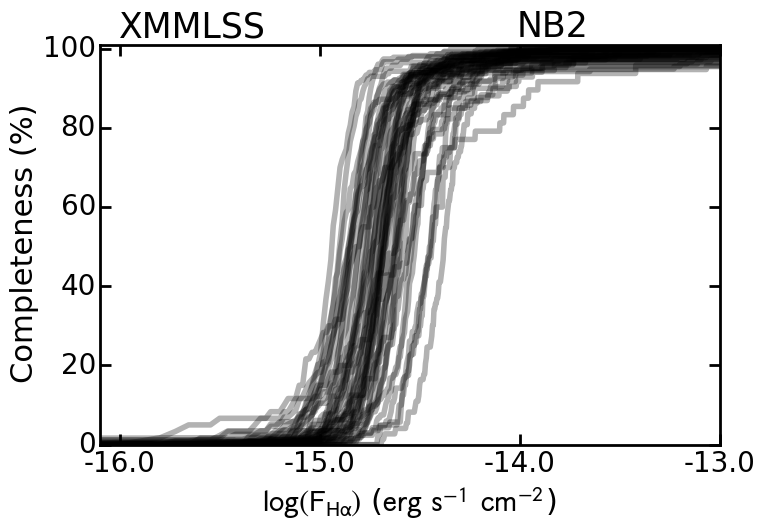}
\end{subfigure}
\caption{Survey completeness as a function of H$\alpha$ flux, plotted separately for each field and NB filter used to select H$\alpha$ candidates. Each curve is associated with the the completeness study for a different CCD chip within each pointing. The darker the colour the more completeness curves fall within that region. Note the XMMLSS field is significantly more complete than the W2 field.}
\label{fig:completeness}
\end{figure*}

The method for studying the completeness is detailed in \S\ref{sec:LHA:completeness}. The dependence of the completeness on line flux is shown in Figure \ref{fig:completeness}.

\section{Survey completeness}
\begin{figure*}
\centering
\begin{subfigure}[b]{0.32\textwidth}
\includegraphics[trim=0cm 0cm 0cm 0cm, width=0.995\textwidth ]{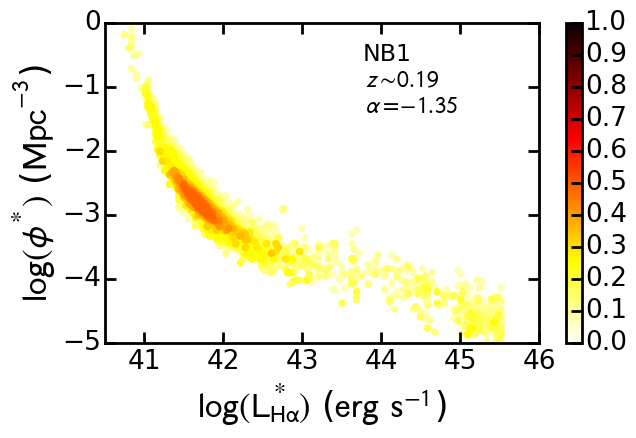}
\end{subfigure}
\hspace{3pt}
\begin{subfigure}[b]{0.32\textwidth}
\includegraphics[trim=0cm 0cm 0cm 0cm, width=0.995\textwidth ]{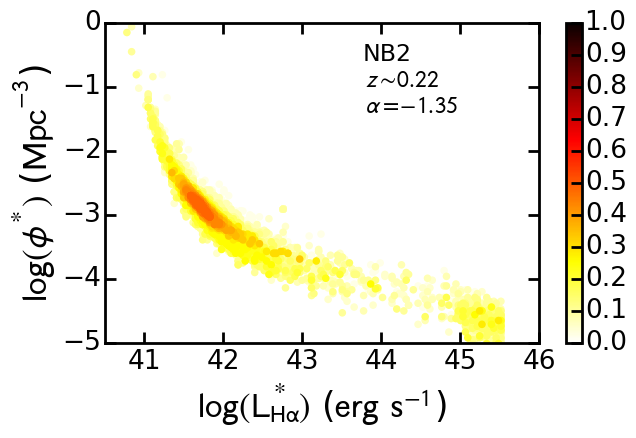}
\end{subfigure}
\hspace{3pt}
\begin{subfigure}[b]{0.32\textwidth}
\includegraphics[trim=0cm 0cm 0cm 0cm, width=0.995\textwidth ]{img/comb_Phi_vs_L_-1.35.png}
\end{subfigure}
\begin{subfigure}[b]{0.32\textwidth}
\includegraphics[trim=0cm 0cm 0cm 0cm, width=0.995\textwidth ]{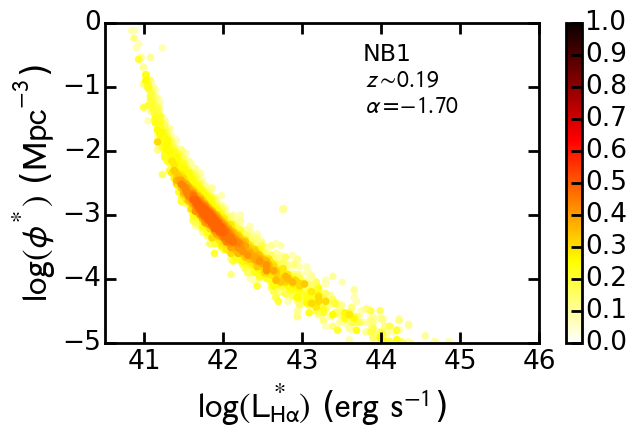}
\end{subfigure}
\hspace{3pt}
\begin{subfigure}[b]{0.32\textwidth}
\includegraphics[trim=0cm 0cm 0cm 0cm, width=0.995\textwidth ]{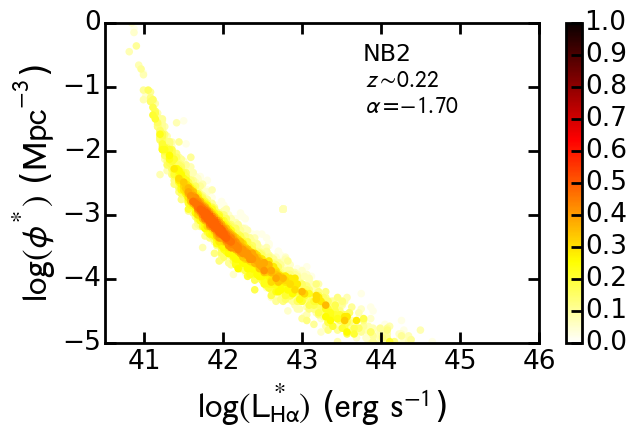}
\end{subfigure}
\hspace{3pt}
\begin{subfigure}[b]{0.32\textwidth}
\includegraphics[trim=0cm 0cm 0cm 0cm, width=0.995\textwidth ]{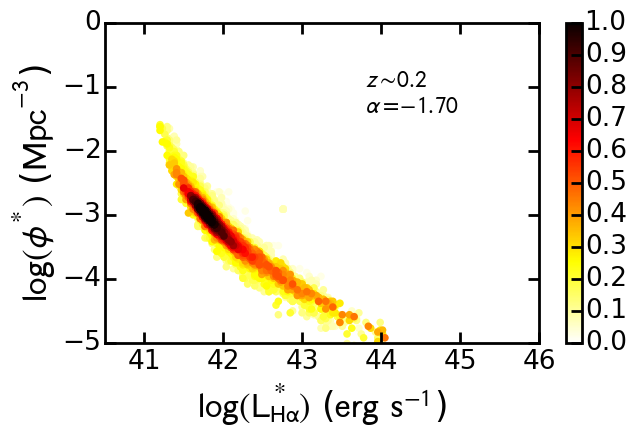}
\end{subfigure}
\caption{As for Figure~\ref{fig:LF}, but with different values of $\alpha$ and when using the data for the two filters separately or together.}
\label{fig:LFall}
\end{figure*}

\begin{figure*}
\centering
\begin{subfigure}[b]{0.49\textwidth}
\includegraphics[trim=0cm 0cm 0cm 0cm, width=0.995\textwidth ]{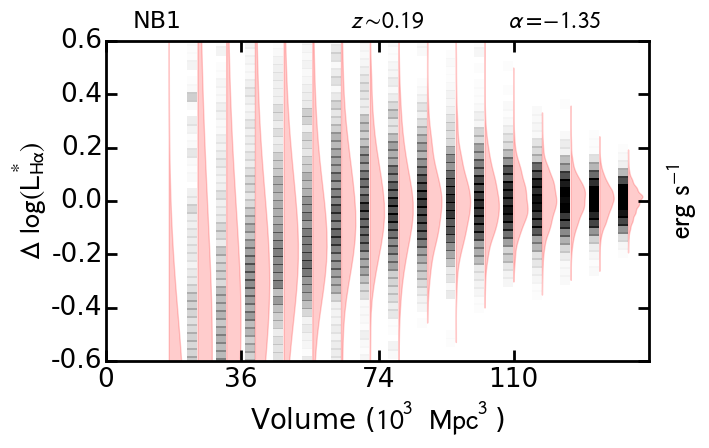}
\end{subfigure}
\begin{subfigure}[b]{0.49\textwidth}
\includegraphics[trim=0cm 0cm 0cm 0cm, width=0.995\textwidth ]{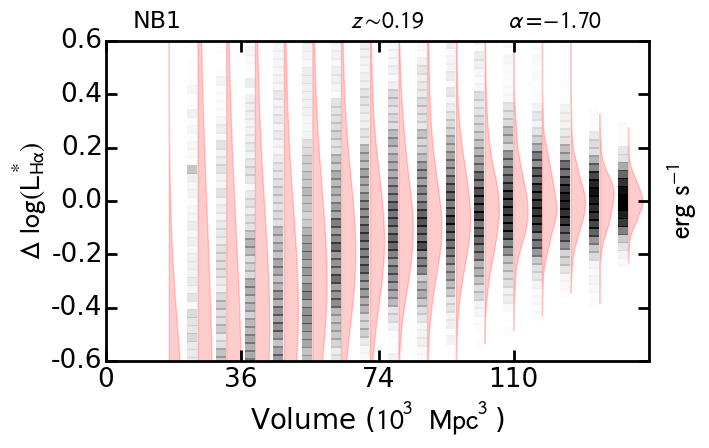}
\end{subfigure}
\hspace{3pt}
\begin{subfigure}[b]{0.49\textwidth}
\includegraphics[trim=0cm 0cm 0cm 0cm, width=0.995\textwidth ]{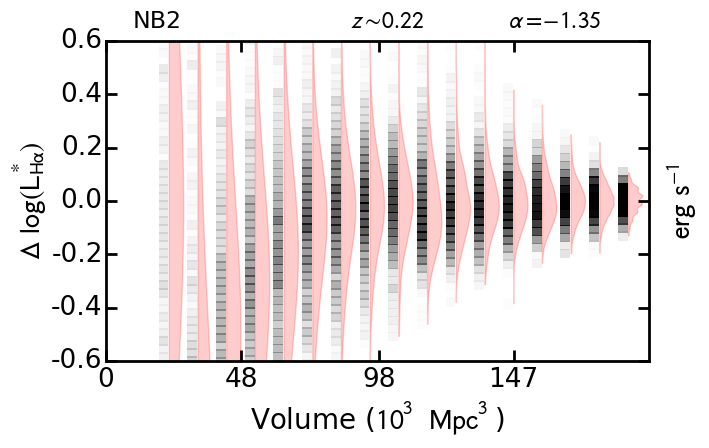}
\end{subfigure}
\hspace{3pt}
\begin{subfigure}[b]{0.49\textwidth}
\includegraphics[trim=0cm 0cm 0cm 0cm, width=0.995\textwidth ]{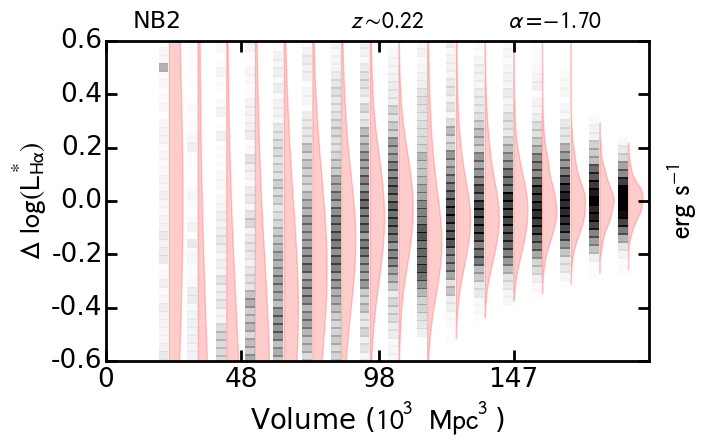}
\end{subfigure}
\caption{As Figure \ref{fig:Lphi}, but for data samples from the two NB filters independently. Note that similar results are found for the two filters, even when considered separately.}
\label{fig:LL}
\end{figure*}

\begin{figure*}
\centering
\begin{subfigure}[b]{0.49\textwidth}
\includegraphics[trim=0cm 0cm 0cm 0cm, width=0.995\textwidth ]{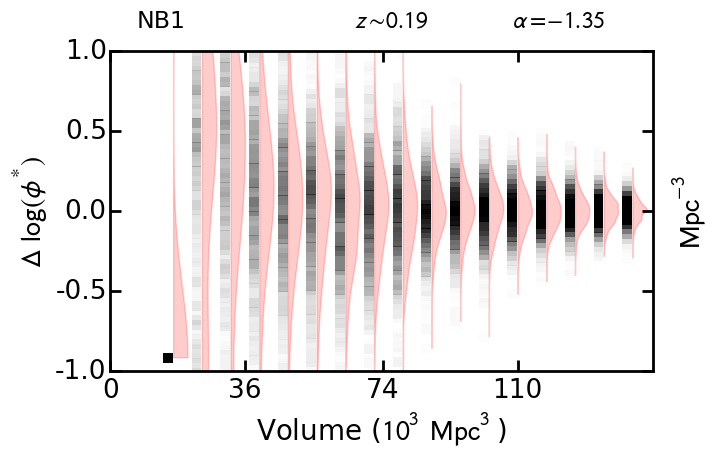}
\end{subfigure}
\begin{subfigure}[b]{0.49\textwidth}
\includegraphics[trim=0cm 0cm 0cm 0cm, width=0.995\textwidth ]{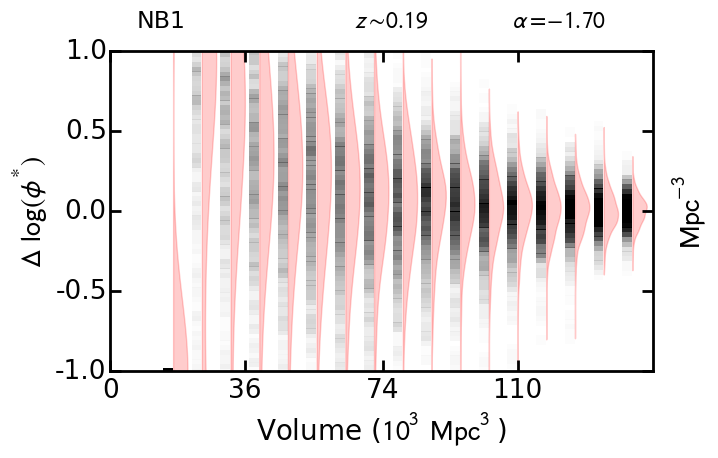}
\end{subfigure}
\hspace{3pt}
\begin{subfigure}[b]{0.49\textwidth}
\includegraphics[trim=0cm 0cm 0cm 0cm, width=0.995\textwidth ]{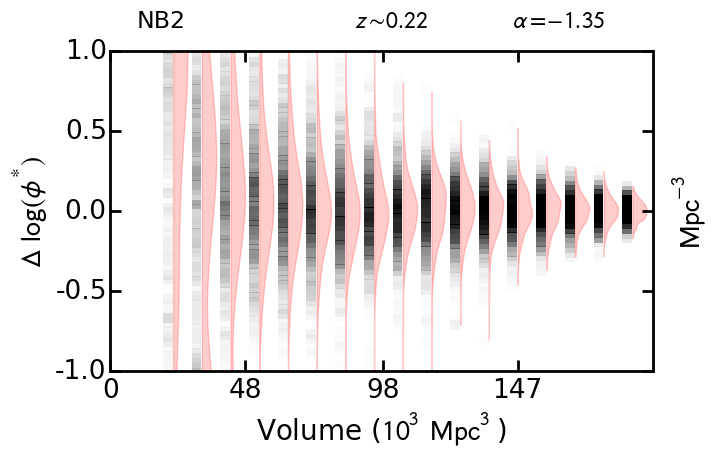}
\end{subfigure}
\hspace{3pt}
\begin{subfigure}[b]{0.49\textwidth}
\includegraphics[trim=0cm 0cm 0cm 0cm, width=0.995\textwidth ]{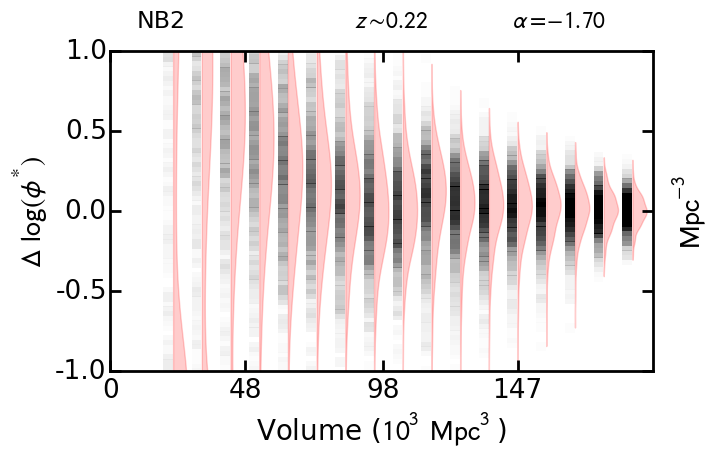}
\end{subfigure}
\caption{As Figure \ref{fig:Lphi}, but for data samples from the two NB filters independently. Note that similar results are found for the two filters, even when considered separately.}
\label{fig:phiphi}
\end{figure*}

\begin{figure*}
\centering
\begin{subfigure}[b]{0.49\textwidth}
\includegraphics[trim=0cm 0cm 0cm 0cm, width=0.995\textwidth]{img/comb_L_distrib_-1.35.png}
\end{subfigure}
\begin{subfigure}[b]{0.49\textwidth}
\includegraphics[trim=0cm 0cm 0cm 0cm, width=0.995\textwidth]{img/comb_Phi_distrib_-1.35.png}
\end{subfigure}
\begin{subfigure}[b]{0.49\textwidth}
\includegraphics[trim=0cm 0cm 0cm 0cm, width=0.995\textwidth]{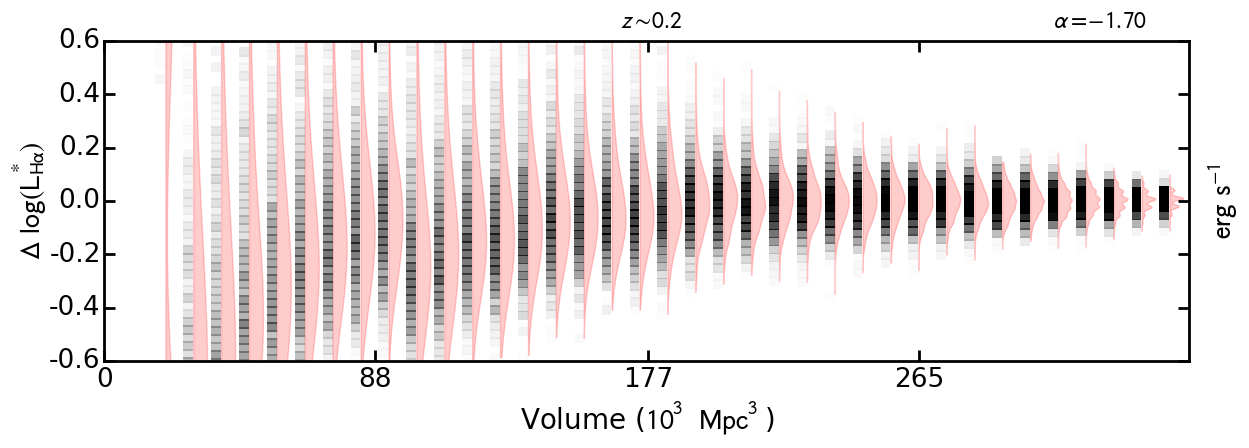}
\end{subfigure}
\begin{subfigure}[b]{0.49\textwidth}
\includegraphics[trim=0cm 0cm 0cm 0cm, width=0.995\textwidth]{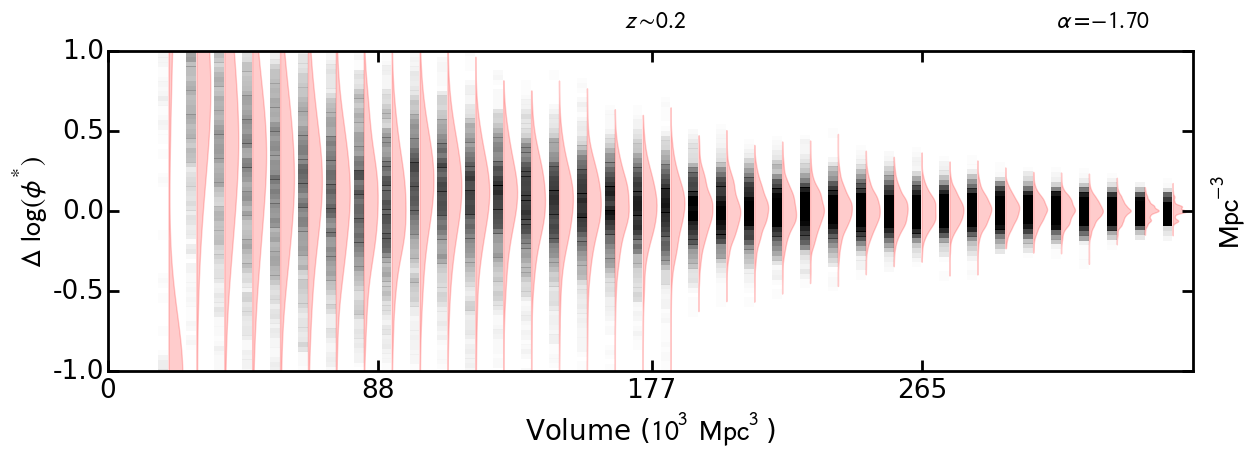}
\end{subfigure}
\caption{As Figure \ref{fig:Lphi}, but for different $\alpha$ values.}
\label{fig:Lphiall}
\end{figure*}

The results of the resampling of the LF at $z\sim0.2$ with different binnings is presented for a range of data selections. The faint end slope is fixed at $-1.35$ and $-1.7$ and $\phi$ and $L$ are fit using data selected from the two NB filters independently and combined. The results are shown in Figures \ref{fig:LFall}, \ref{fig:LL}, \ref{fig:phiphi} and \ref{fig:Lphiall}.

\end{document}